\documentclass[aps,prx,superscriptaddress,showpacs,showkeys,floatflix,reprint, nofootinbib]{revtex4-1}
\usepackage{graphicx}
\usepackage{dcolumn}
\usepackage[pdftex]{color}
\usepackage{appendix}
\usepackage{amsmath}
\usepackage{caption}
\usepackage{subcaption}
\usepackage{bm}
\usepackage{soul}
\definecolor{darkgreen}{rgb}{0.133,0.545,0.133}
\definecolor{orange}{rgb}{1.0,0.76,0.02}

\begin{document}

\title{Unexpected variations in the kinetics of solid solution alloys due to local interactions}

\author{Tanmoy Chakraborty}
\email{tchakra7@jhu.edu}
\affiliation{Department of Materials Science and Engineering, Johns Hopkins University, Baltimore, Maryland 21218,
USA}

\author{Jutta Rogal}
\affiliation{Department of Chemistry, New York University, New York, NY 10003, USA}
\affiliation{Fachbereich Physik, Freie Universit\"{a}t Berlin, 14195 Berlin, Germany }

\date{\today}

\begin{abstract}
Diffusion of atoms in solids is one of the most fundamental kinetic processes that ultimately governs many materials properties. Here, we report on a combined first-principles and kinetic Monte Carlo study of macroscopic diffusion properties of disordered Ti-Ta alloys over the entire composition range. Using simple cluster expansion model Hamiltonians parametrized on density functional theory data, we compute transport properties explicitly including local interactions between the two atomic species and compare them with the non-interacting diffusion model for disordered, random alloys. Surprisingly, we find that although these alloys thermodynamically behave as nearly random solid solutions, their kinetic properties deviate significantly from the behavior predicted by diffusion models for non-interacting systems. We attribute these differences in transport properties to the local interactions that create a rather corrugated potential energy landscape and consequently give rise to energetically non-degenerate end-states of diffusion processes which cannot be realized in a non-interacting disordered or other simpler diffusion models. The findings emphasize the limitations of the widely known non-interacting disordered diffusion model for such systems. Furthermore, we explain that changes in mobility in these alloys is predominantly due to  changes in the correlation factor caused by the local interactions. Our work thus highlights the importance of explicitly including local interactions when assessing the transport properties of thermodynamically nearly disordered alloys.  
\end{abstract}

\maketitle


\section{Introduction}

Solid-state diffusion in multi-component alloys is one of the most fundamental kinetic phenomena that plays a crucial role behind numerous physical processes which ultimately determine many materials properties. These processes include, but are not limited to, solid-state phase transformation~\cite{Christian, PhysRevB.62.203, CAHN1961795, HILLERT19994481, MOORE2002943}, precipitation kinetics~\cite{Clouet, PhysRevB.76.214102}, solute segregation to defects~\cite{Schwarz, Kontis, Hondros, Rice}, charging/discharging in a battery or a fuel cell~\cite{doi:10.1021/ar200329r, C0EE00717J, Puchala}, particular microstructure formation~\cite{MOORE2002943, MaY, Natarajan1}, or shape memory effects~\cite{Pio1}. 
There have been numerous studies on solid-state diffusion based on first-principles calculations of either dilute alloys~\cite{PhysRevLett.100.215901, NAGHAVI2017467, MANTINA20094102, PhysRevB.79.054304, PhysRevB.81.054116, PhysRevB.96.094105, DING2014130, GANESHAN20113214, CHOUDHURY20111, LU2018161, doi:10.1080/14786435.2017.1340685, PhysRevB.94.054106, PhysRevLett.118.105901, doi:10.1080/14786430500228390, doi:10.1080/14786435608238133, doi:10.1080/01418618308243130, doi:10.1080/14786435.2016.1212175} or non-dilute alloys~\cite{doi:10.1080/01418610008212047, doi:10.1080/01418610008212070, doi:10.1080/01418610008212131, doi:10.1080/01418610108216635, PhysRevB.64.184307, PhysRevLett.94.045901, PhysRevB.81.064303, PhysRevB.83.144302, PhysRevMaterials.3.093402, BARNARD2014258, PhysRevMaterials.2.123403, PhysRevLett.121.235901}. A brief, albeit incomplete, list of atomistic methods that are typically being used to estimate diffusion properties in solid-state systems can be primarily categorized into two main groups: ({\it i}) analytical approaches - which include master-equation methods like the self-consistent
mean-field method~\cite{doi:10.1080/01418610008212047, doi:10.1080/14786430500228390} and kinetic mean-field approximations~\cite{Belashchenko_1998, Vaks1, Vaks2}, Green function methods~\cite{doi:10.1080/14786435.2017.1340685, doi:10.1080/01418618308243130, doi:10.1080/14786435.2016.1212175}, five-frequency model~\cite{doi:10.1080/14786435608238133} etc., and ({\it ii}) numerical approaches - which involve either use of kinetic Monte Carlo (kMC)~\cite{PhysRevB.64.184307, PhysRevLett.94.045901, PhysRevB.81.064303, PhysRevB.83.144302, PhysRevMaterials.3.093402, PhysRevMaterials.2.123403, PhysRevB.76.214102, PhysRevMaterials.3.095601} or Monte Carlo (MC) simulations~\cite{PhysRevLett.94.045901, PhysRevLett.121.235901}.

In an atomistic approach, simulating diffusion processes involves the standard treatment of substitutional diffusion which is based on the assumption that atoms migrate via exchange with vacancies (defects) within the crystal with a certain migration barrier. In a non-interacting random alloy, there is no interaction among the involved chemical species which gives rise to \emph{symmetric} diffusion barriers where the potential energy landscape is flat, i.e. the end-states of a diffusion hop are energetically degenerate. This implies that the diffusing species do not have a preference to occupy any particular sites in the system. Simple analytical models based on single migration barriers can be used in this case to calculate transport properties. For non-dilute systems, the situation is more complex since the atomic diffusion processes depend on the local chemical environment, requiring the computation of diffusion barriers for all possible configurations.
To calculate diffusion properties for such systems either analytical models~\cite{Moleko, PhysRevB.4.1111, doi:10.1080/01418618608244021} are used for non-interacting random alloys or MC/kMC are used in combination with density functional theory (DFT) parametrized model Hamiltonians~\cite{PhysRevLett.94.045901, PhysRevMaterials.3.093402, PhysRevB.91.224109, Xi, PhysRevMaterials.2.123403}. The studied systems are primarily either chemically ordered  where strong short- and long-range interactions prevail, or disordered  where \emph{symmetric} diffusion barriers are observed~\cite{PhysRevMaterials.2.123403}. For nearly random alloys that do not show any significant short- and long-range chemical order but still exhibit a {\it rugged} energy landscape with different energies for different configurations, the effect of local interactions on the diffusion properties is less clear. We exemplify this here for Ti-Ta alloys that have recently been suggested a promising materials for high-temperature shape memory applications~\cite{Pio1, yahya}.  In particular, it was shown that the functional properties are strongly composition dependent~\cite{Pio1, Tanmoy-PRB, Tanmoy-JPCM-2021, Alberto-PRM}, and at higher temperature diffusion and phase separation can play an important role in the transformation between different phases and, correspondingly, impact the functional properties.

Specifically, we combine DFT with kMC simulations to estimate transport properties in Ti-Ta  nearly random alloys and compare the results with values predicted by  diffusion models for non-interacting systems. DFT calculations reveal local attractive interactions between Ti and vacancies and Ti-Ti which we incorporate in kMC simulations via simple two-body cluster expansion (CE) Hamiltonians. We find that upon incorporating local interactions, transport properties in these alloys significantly differ from  non-interacting diffusion models, even though the local interactions are too weak to induce any chemical ordering or phase-separation in these alloys in thermodynamic equilibrium. Based on these findings and analysing the results of other, simpler diffusion models, we find that the local interactions in these alloys impact the macroscopic transport properties through a {\it rugged} energy landscape and corresponding {\it asymmetric} diffusion barriers. Our work  highlights the importance of local interactions in estimating transport properties of non-dilute disordered alloys.

The paper is organized as follows: In Section~\ref{diffusion-models}, we discuss interacting and non-interacting diffusion models with corresponding analytical and numerical solutions. Computational details of the DFT and kMC setup are presented in Section~\ref{app:comp-details}. In section~\ref{Results and discussions}, we discuss our results on Ti-Ta alloys, including interaction energy and vacancy formation energy calculations, description of model Hamiltonians for kMC simulations, analysis of short-range order parameter and cluster distribution, transport properties and correlation factors. Finally we summarize and conclude our work in Section~\ref{Summary and conclusion}.


\section{Diffusion models}\label{diffusion-models}
We focus on substitutional diffusion in binary (nearly) random bcc Ti-Ta alloys. In particular, our model represents a perfect bcc crystal, where the number of vacancies is conserved and the presence of extended defects, such as dislocations and grain boundaries that may serve as sources and sinks for vacancies, is neglected. For a binary A-B alloy the diffusion matrix \textbf{D} is related to the Onsager transport coefficients \textbf{L} as~\cite{VANDERVEN201061}
\begin{equation}\label{eq:D}
    \begin{bmatrix} D_{\text{A,A}} & D_{\text{A,B}} \\ D_{\text{B,A}} & D_{\text{B,B}} \end{bmatrix} = \begin{bmatrix} \tilde{L}_{\text{A,A}} & \tilde{L}_{\text{A,B}} \\ \tilde{L}_{\text{B,A}} & \tilde{L}_{\text{B,B}} \end{bmatrix} \begin{bmatrix} \tilde{\Theta}_{\text{A,A}} & \tilde{\Theta}_{\text{A,B}} \\ \tilde{\Theta}_{\text{B,A}} & \tilde{\Theta}_{\text{B,B}} \end{bmatrix} ,
\end{equation}
where $\tilde{L}_{ij}$ = $L_{ij}$ $\Omega$ $k_{\text{B}}$T and $\Omega$ is the volume of a substitutional site. $\tilde{\textit{$\Theta$}}$ represents the thermodynamic factor matrix. The eigenvalues of \textbf{D} provide key physical properties, e.g., the larger eigenvalue $\lambda^{+}$ asymptotically converges to the vacancy diffusion coefficient in the dilute vacancy limit. The smaller eigenvalue $\lambda^{-}$ represents the intermixing diffusion coefficient of the involved species in a binary alloy, i.e., the rate with which the two components A and B of the A-B alloy intermix in the presence of concentration gradients~\cite{VANDERVEN201061, Kehr}. To solve Eq.~\eqref{eq:D}, analytical or numerical approaches can be adopted depending on the complexity of the system, such as whether the system under investigation is a dilute or non-dilute alloy, dependence of diffusion barriers on local environments, exchange frequencies of the involved species with the vacancy etc. We discuss two diffusion models in the following subsections that are of particular interest to our study, namely non-interacting and full-interacting models.

\subsection{Non-interacting model}
We start with the simplest diffusion model which we call the \emph{non-interacting} (NI) model. In the NI model a single diffusion barrier for each species is considered and the migration barrier is independent of the overall composition or local environment. This rather simple diffusion model can be applied to a solid-solution system, where the movement of a diffusing atom is not affected by the other involved chemical species and thus always produces \emph{symmetric} diffusion barrier. To calculate transport properties within the NI model the following approaches can be used. 

\subsubsection{Analytical}\label{NI-ana}
Within the NI model, the analytical equations derived by Moleko~\cite{Moleko} can be used to calculate $L_{\text{AB}}$ for a random binary alloy, where the diffusing species have different exchange frequency, $\Gamma_i$, with which $i$th atomic species exchanges site with a nearest neighbor vacancy ($i$ = A or B), as

\begin{equation}\label{LAA}
    \widetilde{L}_{\text{A,A}} = x_{V}x_{\text{A}}\rho a^{2}\Gamma_{\text{A}}\bigg(1 - \frac{2x_{\text{B}}\Gamma_{\text{A}}}{\Lambda}\bigg) , 
\end{equation}
\begin{equation}\label{LBB}
    \widetilde{L}_{\text{B,B}} = x_{V}x_{\text{B}}\rho a^{2}\Gamma_{\text{B}}\bigg(1 - \frac{2x_{\text{A}}\Gamma_{\text{B}}}{\Lambda}\bigg) ,
\end{equation}
\begin{equation}\label{LAB}
    \widetilde{L}_{\text{A,B}} = \frac{2\rho a^{2}\Gamma_{\text{A}}\Gamma_{\text{B}}x_{V}x_{\text{A}}x_{\text{B}}}{\Lambda} \quad .
\end{equation}
The corresponding expressions for $\Lambda$ and $F$ is given in Appendix~\ref{sec:appA}. $x_V$, $x_{\text{A}}$, and $x_{\text{B}}$ denote the concentrations of vacancy, A, and B, respectively. $a$ represents the hop distance and $\rho$ = $z/2d$, where $z$ and $d$ are the coordination of the each site and dimensionality of the crystal, respectively. $\Gamma_{\text{A}}$ and $\Gamma_{\text{B}}$ represent exchange frequencies of species A and B, respectively, and $\Gamma_{\text{A}}$ $\neq$ $\Gamma_{\text{B}}$. $\Gamma_{i}$ can be approximated within the harmonic approximation in the high-temperature limit as
\begin{equation}\label{gamma-eq1}
\Gamma_{i} = \nu_{0,i} \exp\bigg(-\frac{ E_{i}}{k_{{\text{B}}}T}\bigg) \quad .
\end{equation}
Here, $\nu_{0,i}$ is the attempt frequency (or the vibrational prefactor), $k_{{\text{B}}}$ is the Boltzmann constant, $T$ is the temperature and $E_{i}$ is the migration barrier. One way to obtain $E_{i}$ from first-principles is by employing  combined DFT and nudged elastic band (NEB)~\cite{NEB, doi:10.1063/1.1329672} calculations. The value of $\nu_{0,i}$ can be calculated from harmonic transition-state theory (hTST)~\cite{Pelzer, Eyring, Evans}. 
The thermodynamic factor matrix $\tilde{\Theta}$ in Eq.~\eqref{eq:D} is given by
\begin{equation}\label{theta}
    \tilde{\Theta} = \begin{pmatrix} \frac{(1-x_{\text{B}})}{x_{\text{A}}x_{V}} & \frac{1}{x_{V}} \\ \frac{1}{x_{V}} & \frac{(1-x_{\text{A}})}{x_{\text{B}}x_{V}} \end{pmatrix} \quad .
\end{equation}
Using Eq.~\eqref{LAA}--\eqref{theta}, \textbf{D} can be obtained analytically from Eq.~\eqref{eq:D} within the NI model, and the larger eigenvalue $\lambda^{+}$ can be computed to determine the vacancy diffusivity.

\subsubsection{Numerical}
\label{sec:NInumeric}
The numerical approach to calculate transport properties of binary solid-state systems relies on analyzing trajectories of the involved diffusing species. The macroscopic tracer diffusion coefficient of a given species $i$ can be evaluated using the Green-Kubo formalism~\cite{allnatt1, allnatt2, Allnatt3} from the mean-squared displacement of the atoms given by
\begin{equation}\label{N-D}
    D_i = \lim_{t \to \infty} \frac{1}{6Nt} \sum_{\zeta=1}^{N} \bigg \langle (\Delta \mathbf{R}_{\zeta}^i(t))^2 \bigg \rangle \quad ,
\end{equation}
where $\Delta \mathbf{R}_{\zeta}^{i}$ represents the vector connecting the end-points of the trajectory of atom $\zeta$ of atomic species $i$ ($i$ = A or B for a binary system) after a time interval $t$. $N$ is the total number of atoms of species $i$, $\langle$\ldots$\rangle$ refer to an ensemble average in thermodynamic equilibrium. Likewise, Onsager transport coefficients can be extracted from a trajectory using 
\begin{equation}\label{N-L}
\tilde{L}_{ij} = \frac{\bigg<\bigg(\sum_{\zeta} \Delta \mathbf{R}_{\zeta}^{i} (t)\bigg)\bigg(\sum_{\xi} \Delta \mathbf{R}_{\xi}^{j} 
(t)\bigg)\bigg>}{6tM} \quad ,
\end{equation}
where $M$ represents the number of substitutional sites.    

In general, calculations of diffusion properties by analyzing trajectories of atoms are done using methods such as molecular dynamics (MD). However, solid-state diffusion is a much slower kinetic process, especially at relatively low temperatures, known as {\it rare-events} which cannot be captured using regular MD. In order to simulate diffusion processes over extended time scales, kinetic Monte Carlo~\cite{Bortz, Gillespie} can be used which follows a state-to-state dynamics approach. In a kMC simulation, typically, a lattice model  is employed where possible atomic positions are mapped onto a lattice and the system can evolve by diffusing species hopping between neighboring lattice sites. Thus, solid-state diffusion processes, which are assumed to follow a nearest neighbor hop mechanism, can be replicated using kMC simulations. The rate constant for the exchange of an atom with a neighboring empty site  is one of the key parameters used in kMC simulations, governing the dynamics of the system. The microscopic rate constant, $\Gamma_{i}^{(p)}$, of individual hopping processes, $p$, is again given by Eq.~\eqref{gamma-eq1}, with the corresponding energy barrier $E_i$ as the essential input parameter. For each kMC move, one process $p$ with $\Gamma_{i}^{(p)}$ is chosen among $P$ possible processes for a certain configurations, which in the case of diffusion is equal to the nearest-neighbor sites in a given crystal. The total rate $\Gamma_{\text{tot}}$ is evaluated as the sum over all possible rates for a given configuration, i.e., $\Gamma_{\text{tot}} = \sum_{p}^{P}\Gamma_{i}^{(p)}$. The time step in a kMC simulation is inversely proportional to the total rate, $\Delta t \sim 1/\Gamma_{\text{tot}}$, which generates the correct real time evolution of the system. Time and ensemble averages in Eqs.~\eqref{N-D} and~\eqref{N-L} can be taken as time-weighted averages of different time segments in the kMC simulations, as explained in detail by Du $\emph{et al.}$~\cite{Du}. 

Note, that in a NI model for each species, only a single diffusion barrier entering Eq.~\eqref{gamma-eq1} is required to calculate \textbf{D} and \textbf{L} for both the analytical and numerical approach.
\begin{figure}
\begin{center}
\includegraphics[scale=0.3]{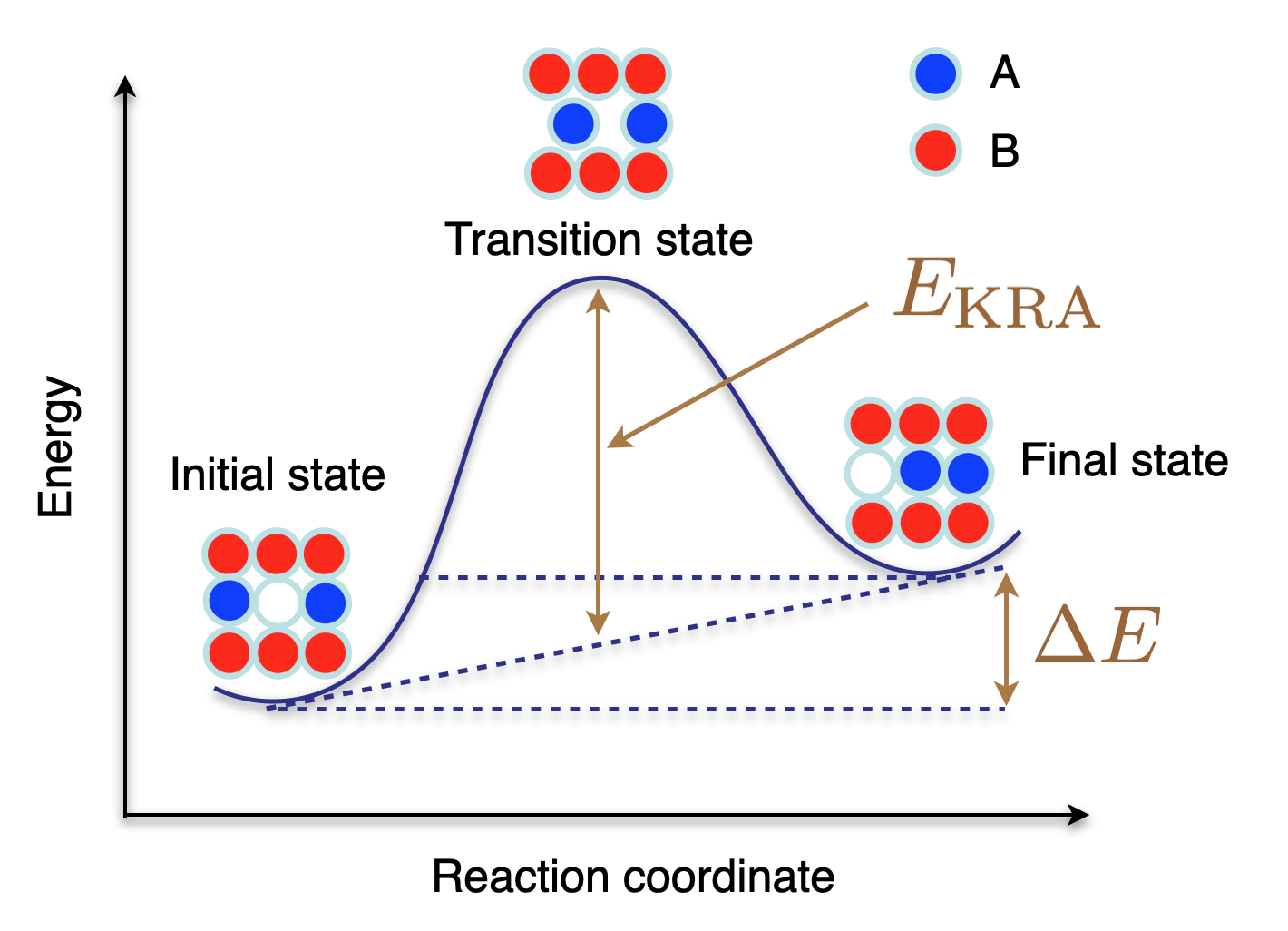}
\caption{\label{KRA-fig}
Schematic representation of a typical asymmetric diffusion barrier in an alloy, where the energy of the initial and final state is different. Blue and red circles represent two different chemical species and the white circle represents the vacancy.}
\end{center}
\end{figure}
\begin{figure}
\begin{center}
\includegraphics[scale=0.25]{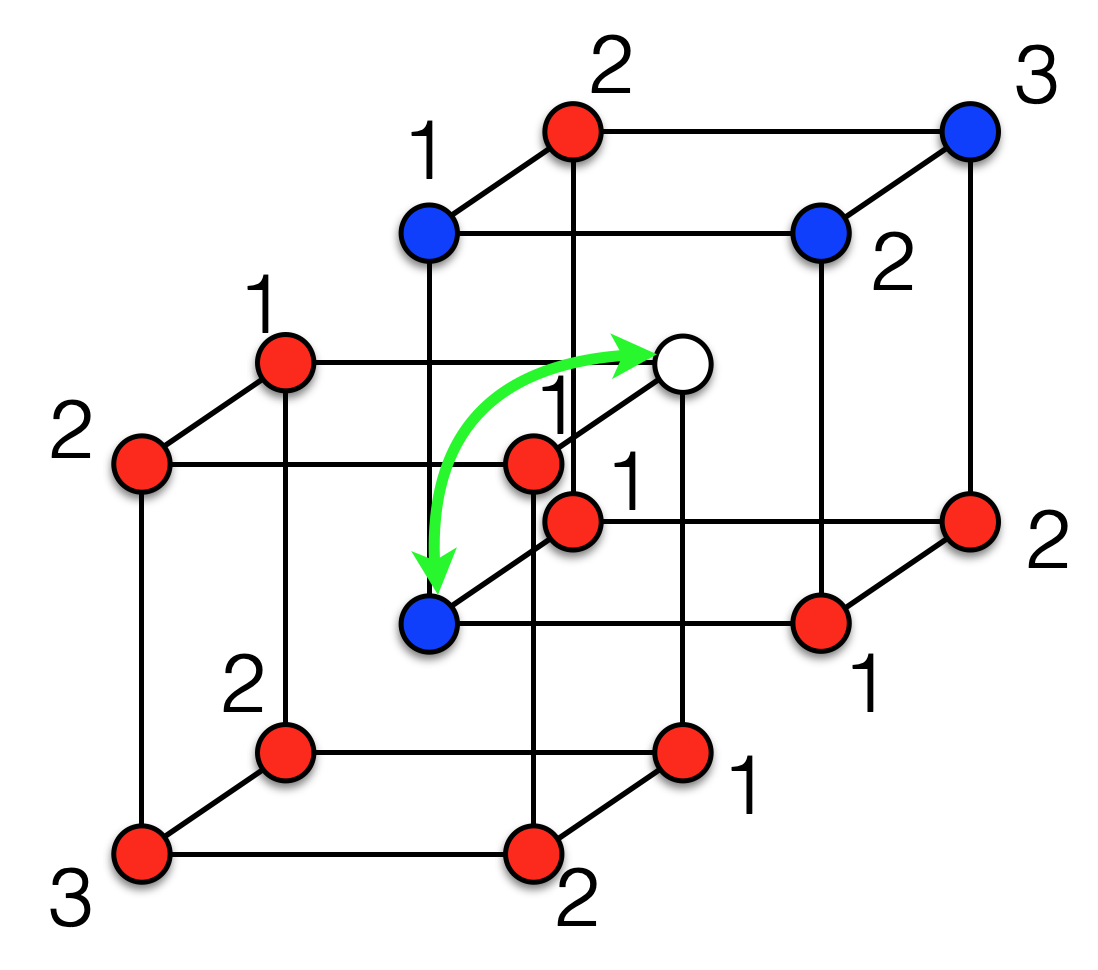}
\caption{\label{diffusion-channel}
Schematic representation of the diffusion channel in a bcc crystal. The picture shows a specific configuration among many possible configurations, where the blue and red circles represent two different chemical species of a binary alloy and the white circle marks the position of a vacancy. The green curve represents the  pathway of the migrating atom. The numbers 1, 2, and 3 denote the first, second, and third nearest neighbor sites of the transition state between the diffusing atom and the vacancy that are within the diffusion channel.}
\end{center}
\end{figure}

\subsection{Full-interacting model}
The full-interacting (FI) model can be used to account for complex interactions of the diffusing species with the local chemical environment surrounding it. Here, $E_{i}$ is both local composition and configuration dependent. This scenario is, in general, more realistic and can capture a {\it rugged} energy landscape, where  $E_{i}$ is $\emph{asymmetric}$ as schematically depicted in Fig.~\ref{KRA-fig}. Due to different chemical environments around an atom before (initial state) and after (final state) the diffusion process, the initial and final state configurations have different energies as a result of local attractive or repulsive interactions. 

An analytical approach to calculate transport properties within the FI model is not available in the literature yet, and would require more elaborate derivations, likely based on the methods discussed in Ref.~\cite{PhysRevLett.121.235901}, where it was shown that transport coefficients can be obtained as minima of local thermodynamically averaged quantities. Computation of all possible diffusion barriers is still required to calculate local thermodynamic averages for a given state and a fast evaluation of averages over many states can be obtained using a CE approach. 
A numerical approach for the FI model, employed in this study, is based again on the Green-Kubo formalism discussed in Section~\ref{sec:NInumeric}. The main  difference is the evaluation of the diffusion barriers $E_{i}$ in Eq.~\eqref{gamma-eq1}, which is  discussed in detail in the following. 

For the diffusion process shown in Fig.~\ref{KRA-fig}, the diffusion barrier is direction dependent, which is rather cumbersome to incorporate into a numerical model.  To circumvent this problem, the diffusion barrier is decomposed into two parts.
The first one accounts for the description of the energy difference between the end-states, $\Delta E$, of the diffusion process and the second one is known as so called kinetically resolved activation (KRA) barrier. The two contributions are given by~\cite{KRA}
\begin{equation}\label{eq:DE-method}
    \Delta E = E_{I} - E_{F} \quad ,
\end{equation}
and
\begin{equation}\label{eq:KRA}
 E_{\text{KRA}} = E^{\text{TS}} - \frac{1}{2} (E_{I} + E_{F}) \quad ,
\end{equation}
where $E_{I}$, $E_{F}$, and $E^{\text{TS}}$ represent the energy of the initial, final, and transition state, respectively. Using Eqs.~\eqref{eq:DE-method} and~\eqref{eq:KRA}, $E_{i}$ can be estimated for a random alloy within the FI model as
\begin{equation}\label{eq:KRA+DE}
    E_{i} = E_{\text{KRA}} - \frac{1}{2} \Delta E \quad .
\end{equation}
On-the-fly numerical evaluation of $\Delta E$ and $E_{\text{KRA}}$ for every diffusion jump during a kMC simulation using a combined DFT and NEB approach  is computationally unfeasible. Moreover, there are too many configurations to enumerate all possibilities beforehand and store the results in a database. Alternatively,  a model Hamiltonian can be constructed that is parametrized using DFT and NEB data and can then  be used directly in kMC simulations to study the time evolution of the system. An efficient way to do this, is to use a cluster expansion (CE) approach~\cite{CE1, CE2} which provides a fast and accurate evaluation of configurational energies and  migration barriers. Finally, to solve Eq.~\eqref{eq:KRA+DE}, construction of two CE models is required: (1) configurational CE to predict $\Delta E$, $\Delta E_{\text{config-CE}}$ by taking the differences between two configurational energies and (2) a local CE to predict $E_{\text{KRA}}$, $E_{\text{KRA-CE}}$ (cf. Appendix~\ref{CE-formalism} for CE formalism).


\section{Computational details}\label{app:comp-details}
\subsection{DFT calculations}
DFT calculations were carried out for bcc Ti-Ta alloys using the projector augmented wave (PAW)~\cite{Bloechl} method as implemented in the Vienna $Ab$ $initio$ Simulation Package (VASP)~\cite{Kresse_15, Kresse_11169, Kresse_1758}. 
The PAW potentials include $3p$ and $5p$ electrons in the valence shell for Ti and Ta, respectively. The  Perdew-Burke-Ernzerhof parametrization of the generalized gradient approximation was used for the exchange-correlation functional (PBE-GGA)~\cite{Perdew_3865}.
The calculations were performed with an energy cutoff of 300~eV for the plane waves and the Methfessel-Paxton scheme was used to integrate the Brillouin zone (BZ) with a smearing of $\sigma$ = 0.05 eV. Within our computational setup, total energies were converged to within 1 meV/atom. Structures were relaxed until all forces were below $10^{-3}$~eV/\AA. For the Ti-Ti interaction energy, we use a 3$\times$3$\times$3 bcc supercell containing 54 atoms. A 7$\times$7$\times$7 $k$-point mesh is used. The cell shape and volume are kept fixed during the calculations to that of bcc Ta, however ionic relaxation is performed until all forces are below 0.01 eV/\AA. The vacancy formation energy calculations were performed using 128-atoms (4$\times$4$\times$4 conventional bcc cells) special quasi-random structures (SQS)~\cite{Zunger} supercells with 4$\times$4$\times$4 $k$-point mesh. To generate the SQS configurations we used a modified version~\cite{Pezold, Kossmann} of the ATAT package~\cite{Walle}. The diffusion barriers are calculated using the NEB method~\cite{NEB, doi:10.1063/1.1329672} as implemented in the VTST package~\cite{vtst} for VASP. For NEB calculations, we use the same cell size and $k$-point mesh as for the interaction energy calculations, i.e a 54-atoms cells. Only nearest neighbor diffusion processes are considered in this study (see Fig.~\ref{diffusion-channel}), thus the energy landscape is sufficiently simple that a single NEB image is sufficient. We have checked that additional images do not change the results. 

\subsection{kMC setup}
For all kMC simulations we use 16$\times$16$\times$16 conventional bcc cells containing 8192 sites and a single vacancy with periodic boundary conditions. The vacancy concentration is thus dilute which allows us to neglect vacancy-vacancy interactions.
The value of $\nu_{0,p}$ is set to $10^{13}$ s$^{-1}$ in all our kMC simulations which is a reasonable approximation for typical solid-state systems. To represent a random alloy system we randomly distribute Ti and Ta atoms in the simulation cell. The jump distance was set to $a$ = 2.87 \AA~which was obtained from the equilibrium lattice constant of pure bcc Ta calculated using DFT.  We run  1$\times$10$^7$ kMC steps for initial equilibration. 

\section{Results and discussions}\label{Results and discussions}
\subsection{Interaction and vacancy formation energy}
\begin{figure}
\begin{center}
\includegraphics[scale=0.25]{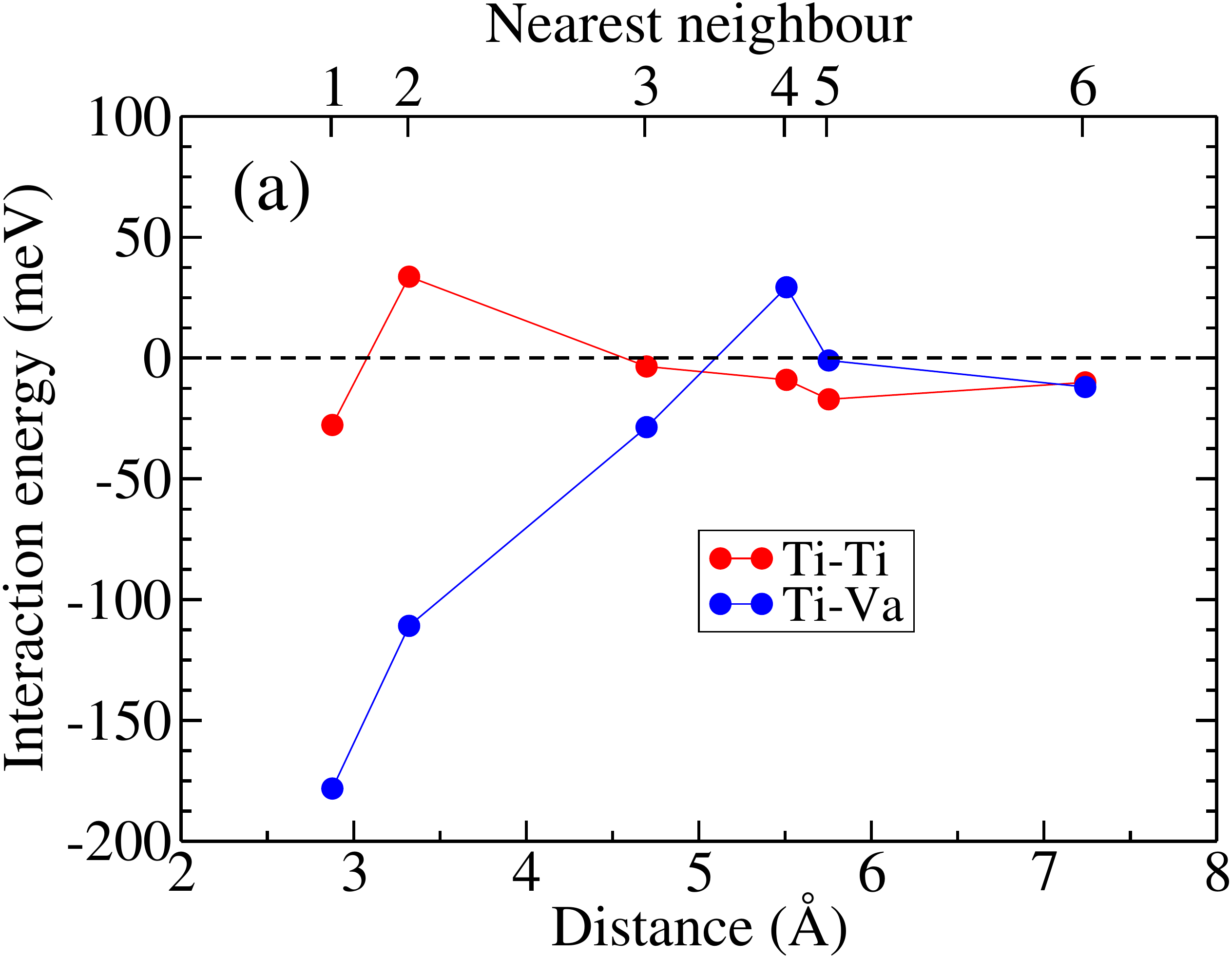}
\includegraphics[scale=0.25]{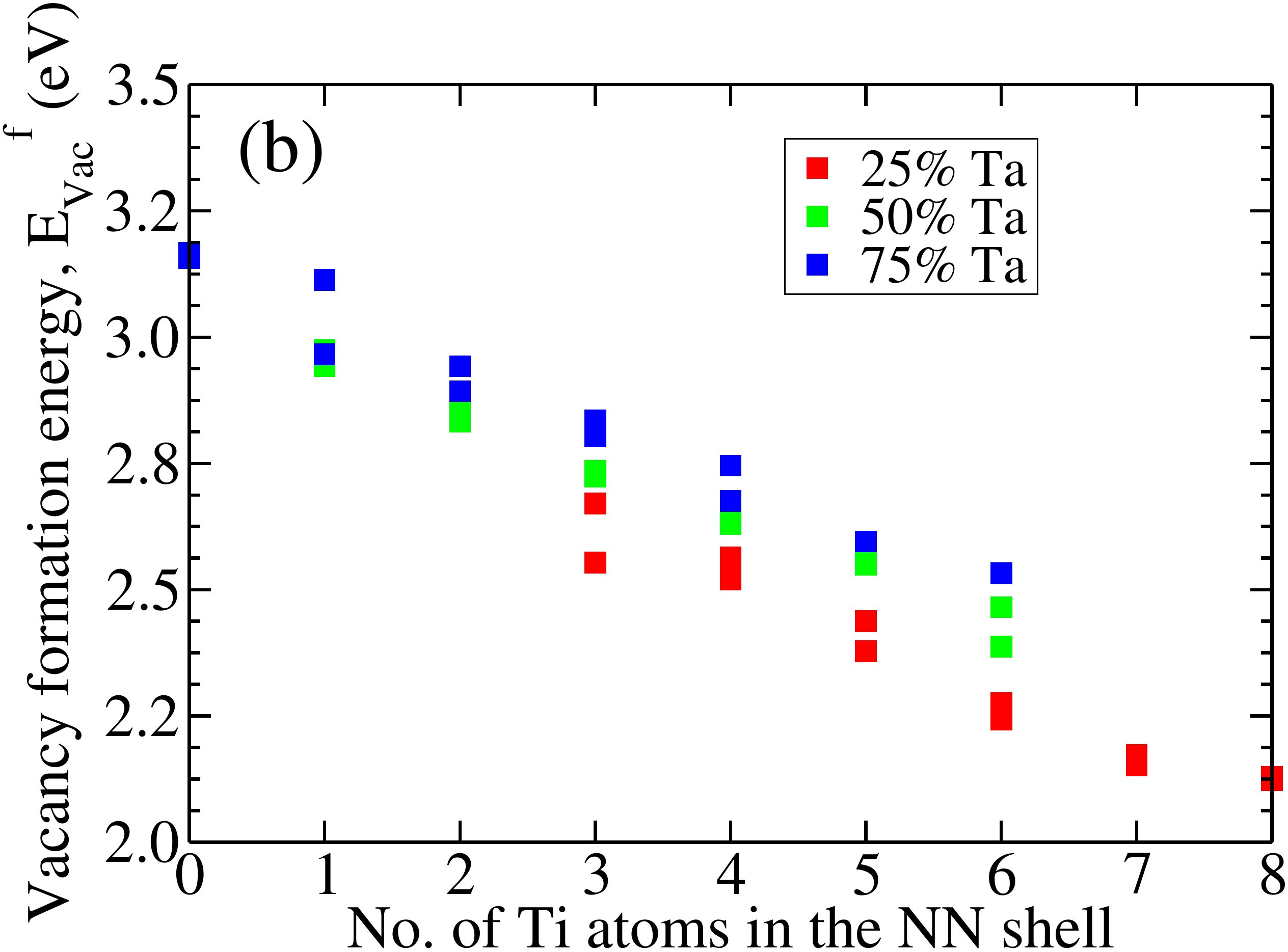}
\caption{\label{fig-3}
(a) DFT calculated Ti-Ti (red) and Ti-Va (blue) interaction energy in bcc Ta as a function of distance. (b) Vacancy formation energy as a function of number of Ti atoms in the nearest neighbor (NN) shell of the vacancy. The red, green, and blue color represent the 25\%, 50\%, and 75\% Ta content case, respectively.}
\end{center}
\end{figure}
The interaction energy is an important entity that acts as a driving force which facilitates the formation of  specific states in an alloy, such as chemically ordered, phase-separated, or solid-solution phases. The Ti-Ti and Ti-Va interaction energies are calculated according to
\begin{equation}\label{eq:Ti-Ti}
 E_{\text{int (Ti-Ti)}} = (E_{\text{Ta}_{52}\text{Ti}_{2}} + E_{\text{Ta}_{54}}) -(2 \times E_{\text{Ta}_{53}\text{Ti}_{1}}) \quad .
\end{equation}
and,
\begin{equation}\label{eq:V-Ti}
 E_{\text{int (Ti-Va)}} = (E_{\text{Ta}_{52}\text{Ti}_{1}} + E_{\text{Ta}_{54}}) - (E_{\text{Ta}_{53}\text{Ti}_{1}} + E_{\text{Ta}_{53}})  \quad .
\end{equation}
$E_{\text{Ta}_{52}\text{Ti}_{2}}$ is the total energy of a supercell containing 52 Ta and 2 Ti atoms. The 2 Ti atoms are placed at various distances relative to each other to obtain the corresponding interaction energies.
$E_{\text{Ta}_{54}}$ represents the total energy of pure bcc Ta with 54 atoms, and $E_{\text{Ta}_{53}\text{Ti}_{1}}$ is the total energy of the supercell containing 1 Ti atom and 53 Ta atoms. According to the definition in Eq.~\eqref{eq:Ti-Ti}, if the term $E_{\text{int}}$ becomes negative, Ti atoms attract each other, while a positive value indicates a repulsive interaction between  Ti atoms. In Fig.~\ref{fig-3}(a) we show the DFT calculated results of Ti-Ti and Ti-vacancy (Ti-Va) interaction energies as a function of distance. We observe an overall attractive Ti-Ti and Ti-Va interaction energy in the first shell, with $\sim$ 30 meV and $\sim$ 175 meV, respectively. In the second neighbor shell, the Ti-Ti interaction becomes repulsive, while the Ti-Va interaction remains attractive. This suggests that Ti atoms do prefer the vicinity of vacancies, but there is also a weak tendency to form Ti clusters.

The vacancy formation energy is one of the key parameters that provides useful insight related to transport properties of a system.  The vacancy formation energy is needed to calculate vacancy concentrations which are again crucial to determine the diffusivity in substitutional diffusion. We calculate the vacancy formation energy, $E_{\text{Vac}}^f$, for a random alloy with different sites, as~\cite{jutta_pssb}
\begin{equation}\label{final_vfe}
 E_{\text{Vac}}^f (\alpha; x_{\text{A}}, x_{\text{B}}) = \frac{1}{N_{\alpha}}\sum_{i_{\alpha} = 1}^{N_{\alpha}}[E_{\text{1Vac}, i_{\alpha}} + \mu_i - E_{\text{perfect}}] \quad.
\end{equation}
In Eq.~\eqref{final_vfe}, the effective interactions between vacancies and involved species $i$ ($i$ = Ti or Ta atoms) are limited to the nearest neighbors only, where $\mu_i$ is the chemical potential of species $i$ (Ti or Ta), $\alpha$ represents a particular neighborhood of a vacancy, and $E_{\text{perfect}}$ represents the energy of a defect-free supercell. $E_{\text{1Vac}, i_{\alpha}}$ signifies the energy of a supercell containing one vacancy created at a site $i_{\alpha}$ with a particular neighborhood. The calculated $E_{\text{Vac}}^f$ of the bcc phase for different neighborhoods (chemical environments) of the vacancy in Ti-Ta alloys is shown in Fig.~\ref{fig-3}(b). Here, we use SQS supercells to mimic disordered bcc Ti-Ta alloys which allows us to realize different chemical neighborhoods of the vacancy. We find that the larger the number of Ti atoms in the nearest neighbor shell of the vacancy, the smaller is the vacancy formation energy. This is consistent with the smaller $E_{\text{Vac}}^f$ in pure hcp Ti as compared to pure Ta. The overall composition has only a minor effect on  $E_{\text{Vac}}^f$; for a given number of Ti atoms in the vicinity of the vacancy, smaller Ti concentrations are associated with higher vacancy formation energies and vice versa. Overall, the vacancy formation energy depends much stronger on the local than the global composition.  In a random alloy, the average local composition will reflect the global composition, and thus we infer a linear change in the vacancy formation energy for Ti-Ta alloys as a function of composition.

\subsection{Hamiltonian construction for the FI model}
\begin{figure}
\begin{center}
\includegraphics[scale=0.35]{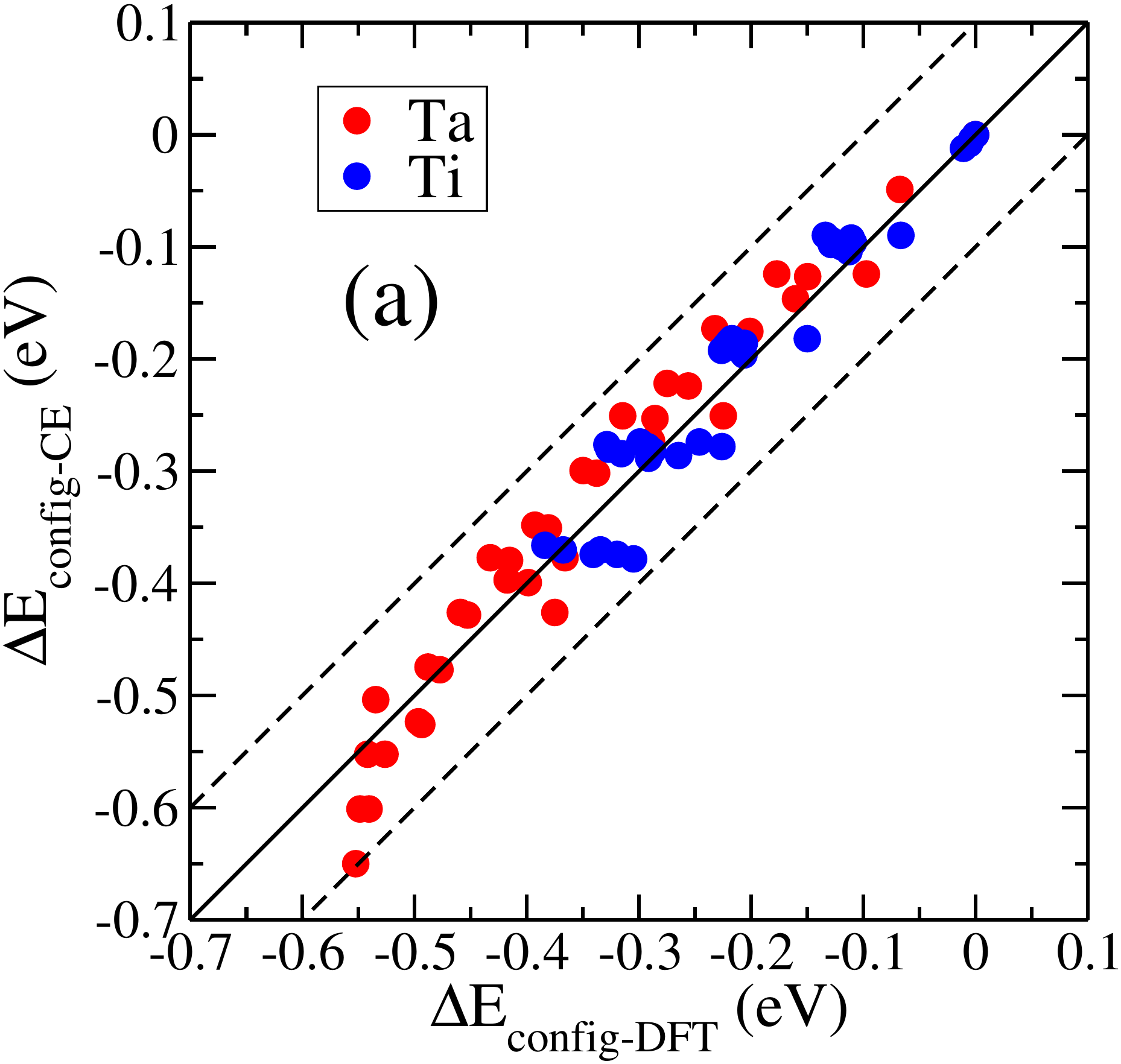}
\includegraphics[scale=0.35]{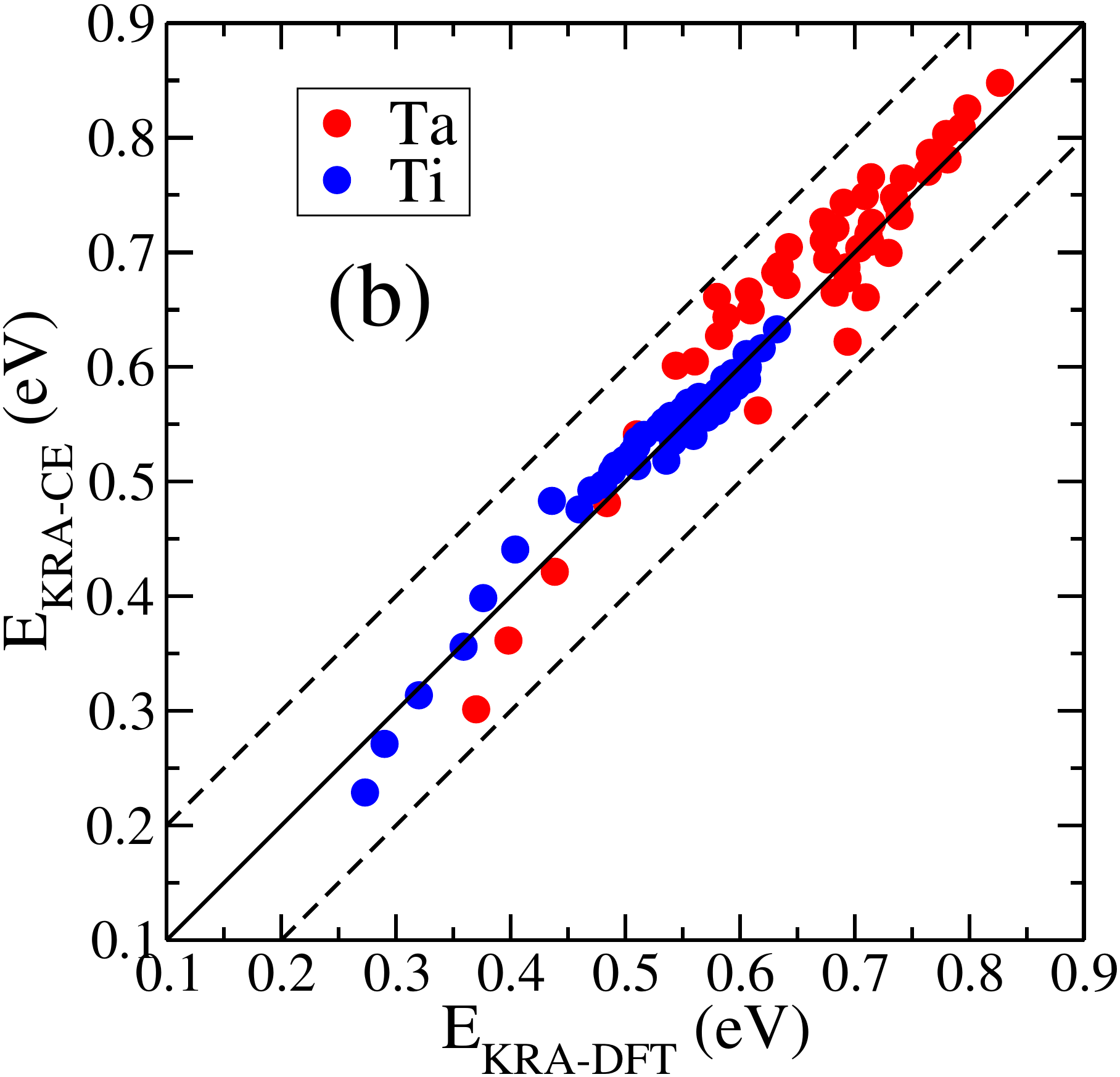}
\caption{\label{KRA_results}
(a) Comparison of DFT calculated end-state energy differences and the CE fitted model. (b) Comparison of the DFT calculated KRA barriers and the CE fitted model. The blue and red color represent the data when the diffusing species is Ti and Ta, respectively.}
\end{center}
\end{figure}
We discuss here the strategy for constructing the Hamiltonian for the FI model. We use a simplified CE approach, which is based on a conventional full CE concept, where we focus on only a few selective pair clusters  within the so-called {\it diffusion channel}. A schematic diffusion channel is shown in Fig.~\ref{diffusion-channel}, which is essentially a nearest neighbor model, i.e, it accounts for the interactions coming from all  sites that are  nearest neighbor of either the migrating species or the vacancy. This implies that the diffusion barriers $E_{i}$ are only affected by the different arrangements of the chemical species  within this diffusion channel, which is motivated by the fact that  the chemical environment in the vicinity of the diffusion path should have the largest impact on the diffusion barrier. This can be considered as a first-order effect on the transition state energy. Similar approaches have been discussed in literature for fcc systems~\cite{Goiri, PhysRevMaterials.2.123403, LEITNER20101091}. 

The total number of possible configurations, due to the different arrangements of Ti and Ta in the diffusion channel, would be nearly 40 million without symmetry taken into account. Even with considering symmetry, the total number of configurations would be too large to consider all of them for DFT calculations. Thus, we apply  a heuristic approach to sample various occupations of the diffusion channel with Ti and Ta, which is based on a systematic selection of different configurations for each composition. Based on this approach, we considered in total 76 symmetrically distinct configurations (38 each for Ti and Ta migration) which comprises of 1, 3, 4, 8, 6, 5, 3 configurations containing 1, 2, 3, 4, 5 and 6 Ti atoms, respectively with different decorations of the diffusion channel. For configurations containing 7 to 14 Ti atoms, we only consider one structure for each case. For each configuration, we compute the diffusion barrier using  NEB+DFT and decompose it into KRA and $\Delta E$, yielding 76 data points. We find that the KRA and $\Delta E$ change significantly depending on the local chemical environment. To incorporate this dependence in our kMC simulations, we develop two simple CE models: (1) a configurational CE to predict  $\Delta E$ and (2) a local KRA-CE to predict the KRA part. Once we build these two CE models, we use Eq.~\eqref{eq:KRA+DE} in our kMC simulations to compute migration barrier for any diffusion hop.

\subsubsection{Configurational cluster expansion} 
The training set for the configurational CE consists of 76 data points for $\Delta E$ (152 DFT configurational energies) which is shown in Fig.~\ref{KRA_results}(a). Here, we divide the data set into two categories: (1) blue symbols represent, when Ti is the diffusing atom, and (2) red symbols, when Ta is the diffusing atom. We find that including pair clusters for Ti-Va and Ti-Ti up to third  and second nearest neighbor, respectively, is reasonable enough to correctly capture the migration barriers. Our expression for the configurational CE model to predict $\Delta E$, resulting from a diffusion jump, is
\begin{multline}\label{eq:DE}
 \Delta E = -[0.1241 \times \Delta n\text{Ti-Va}^{(1\text{NN})}] -  [0.0753 \times \Delta n\text{Ti-Va}^{(2\text{NN})}] \\ +  
 [0.0023 \times \Delta n\text{Ti-Va}^{(3\text{NN})}]
                            - [0.0343 \times \Delta n\text{Ti-Ti}^{(1\text{NN})}] \\ + [0.0103 \times \Delta n\text{Ti-Ti}^{(2\text{NN})}] \quad .
\end{multline}
Here, $\Delta n$ in the first, second, and third term represents the change in Ti-Va pairs in the first, second, and third nearest-neighbor, respectively, which are within the diffusion channel. Likewise, in the fourth and fifth term $\Delta n$ denotes the change in Ti-Ti pairs in the first and second neighbor, respectively, which are within the diffusion channel. The coefficients in Eq.~\eqref{eq:DE} are obtained by fitting the CE expression to the DFT $\Delta E$ data. 
The signs of the coefficients in Eq.~\eqref{eq:DE} suggest that there is an attractive Ti-Va pair interaction for the first and second nearest neighbor, respectively, and a repulsive interaction for the third nearest neighbor. The Ti-Ti pair interaction is attractive for the first nearest neighbor and repulsive for the second nearest neighbor. This is in good agreement  with the DFT data in Fig.~\ref{fig-3}. 
Furthermore, the $\Delta E$ values predicted by our CE model are in overall agreement with the corresponding DFT data, shown in Fig.~\ref{KRA_results}(a).
The leave-one-out cross validation (LOOCV) score for the $\Delta E$ CE is $\sim$ 30 meV.

\subsubsection{KRA cluster expansion}
For the KRA-CE, we have two separate equations, depending on  if Ti or Ta is diffusing:
\begin{multline}\label{Ti_fit}
E^{\text{Ti}}_{\text{KRA}} = 0.568 + (0.0216 \times n\text{Ti}^{(1\text{NN})}) - (0.0166 \times n\text{Ti}^{(2\text{NN})})
 \\ - (0.0424 \times n\text{Ti}^{(3\text{NN})}).
\end{multline}
and
\begin{multline}\label{Ta_fit}
E^{\text{Ta}}_{\text{KRA}} = 0.781 + (0.0222 \times n\text{Ti}^{(1\text{NN})}) - (0.0387 \times n\text{Ti}^{(2\text{NN})}) 
 \\ - (0.06 \times n\text{Ti}^{(3\text{NN})}).
\end{multline}
Again, the coefficients in Eqs.~\eqref{Ti_fit} and~\eqref{Ta_fit} are obtained by fitting the KRA-CE expressions to the KRA-DFT data. $n$ represents the number of Ti atoms that are first, second, and third nearest neighbours to the TS, 
not only within but also outside of the diffusion channel. All  first and second nearest neighbor sites of the TS are within the diffusion channel, however most of the third nearest neighbor sites of the TS are outside the diffusion channel (except two sites, marked by \textbf{3} in Fig.~\ref{diffusion-channel}). The sites that are equidistant from the TS will have equal contribution in affecting the diffusion barriers. Thus, also the third nearest neighbor sites of the TS which are outside of the diffusion channel need to be taken into account for the KRA-CE. The KRAs predicted by our CE are in very good agreement with the DFT data, shown in Fig.~\ref{KRA_results}(b). The LOOCV scores are $\sim$ 16 meV and $\sim$ 35 meV for Ti and Ta KRA-CE models, respectively, which is reasonably accurate and comparable to other studies~\cite{Goiri}.

\subsection{Short range order parameter and cluster distribution}
\begin{figure*}
\begin{center}
\includegraphics[scale=0.22]{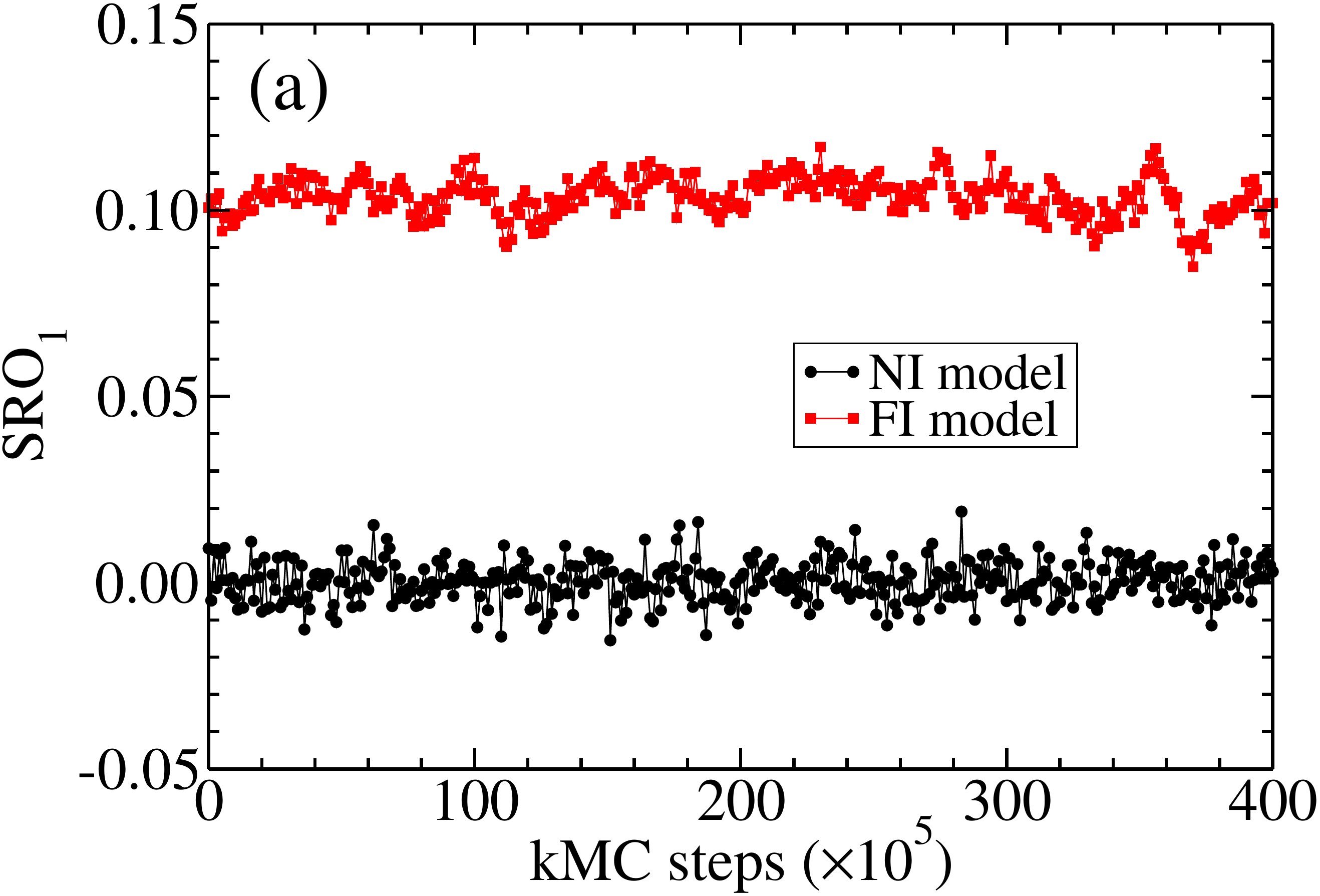}
\includegraphics[height=4.12cm]{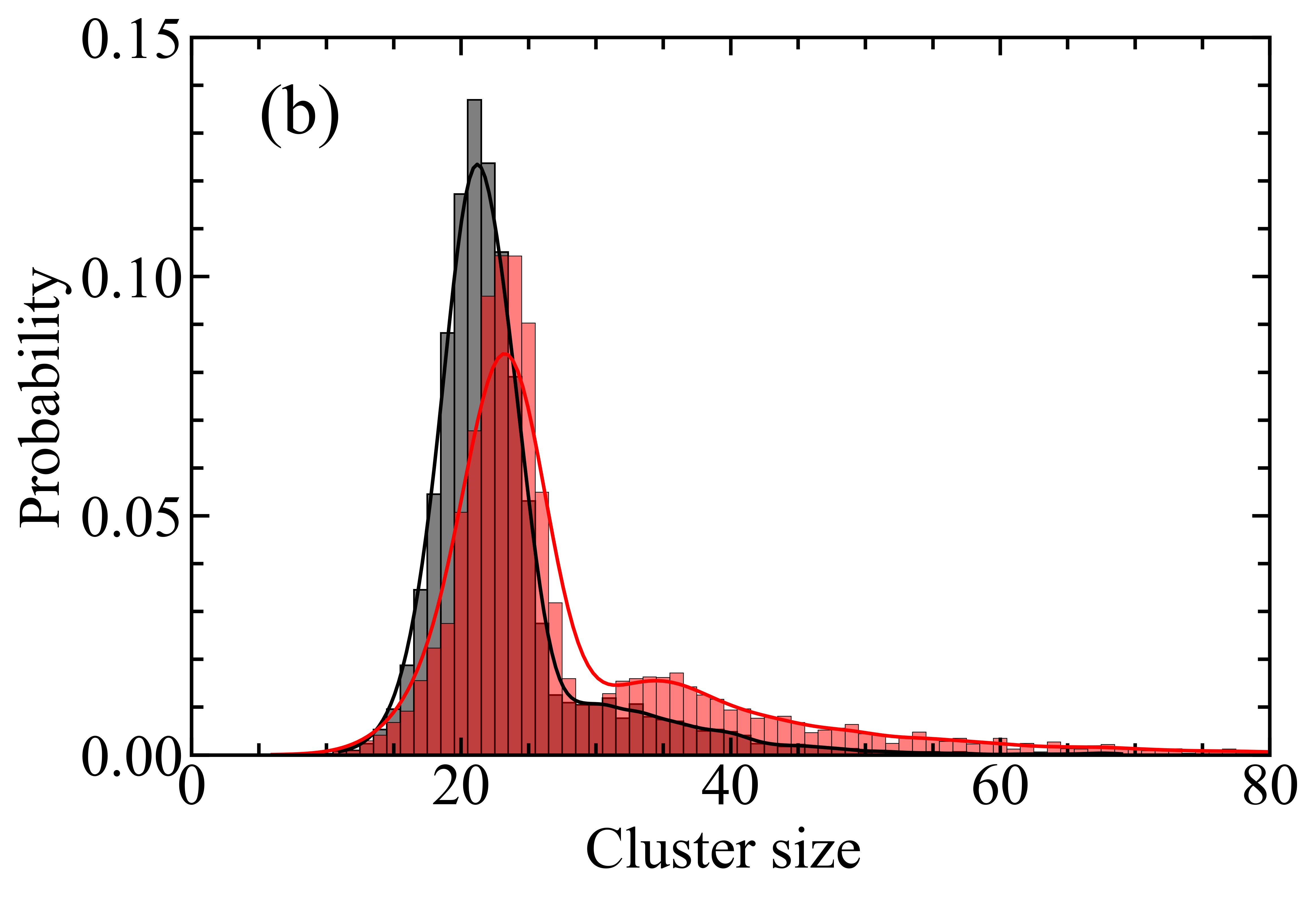}
\includegraphics[scale=0.22]{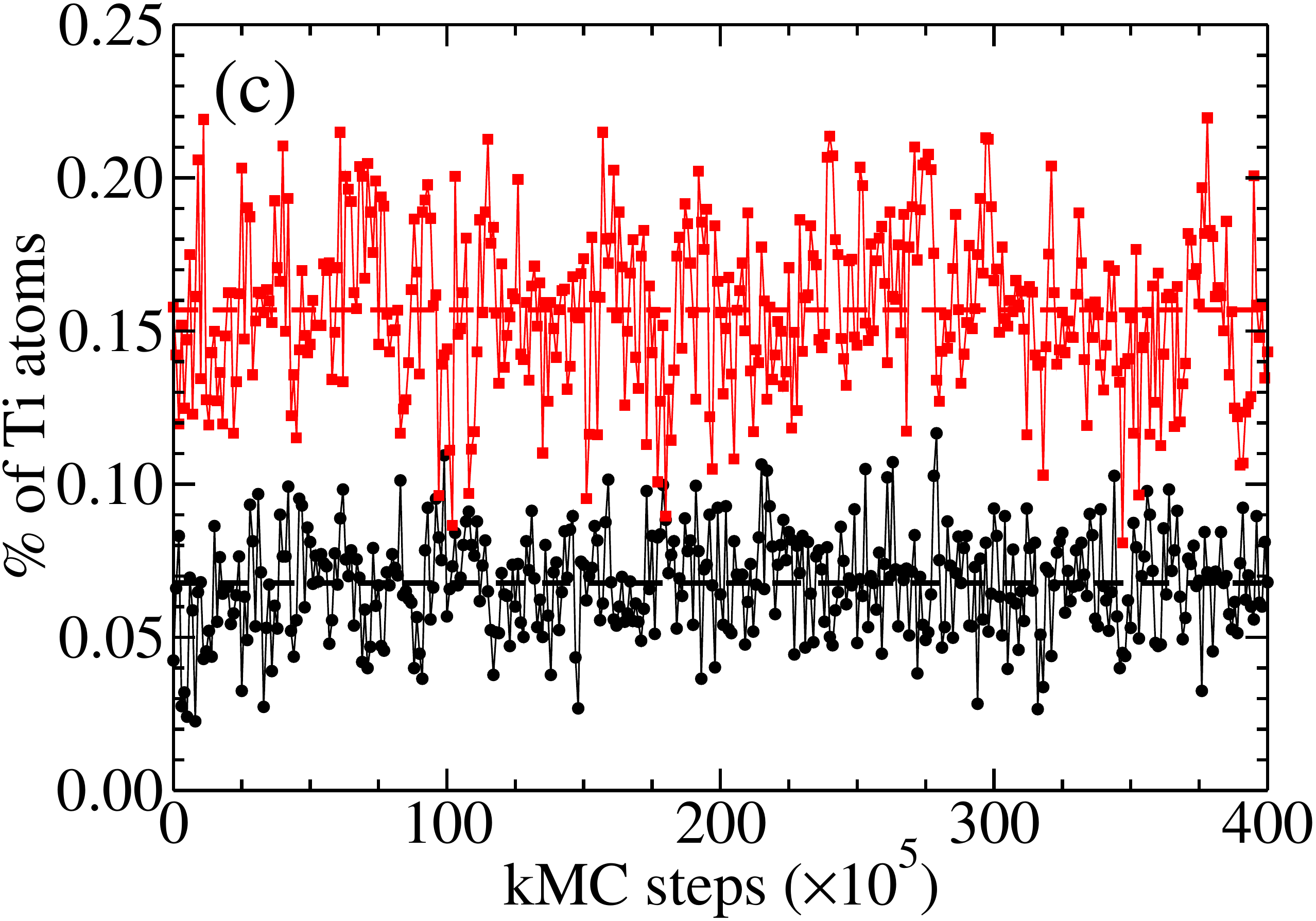}
\caption{\label{fig-5}
Comparison of (a) nearest neighbor SRO parameter (b) cluster size distribution and (c) percentage of Ti atoms participating in cluster formation in (b) calculated using NI (black) and FI (red) diffusion models. All values are computed at $T$ = 900 K and at 50\% Ta content. A total 4$\times$10$^7$ kMC steps were run after 1$\times$10$^7$ initial equilibration.  Values were taken after every 1$\times$10$^5$ kMC steps to avoid correlations.}
\end{center}
\end{figure*}
Ti-Ta alloys are known to be solid solutions at elevated temperatures~\cite{Murray-expt, Murray-theo, BARZILAI2016255} over the entire composition range. 
In order to assess if our CE model including local interactions can reproduce this behaviour or if the system exhibits any chemical ordering/clustering/phase separation phenomena, we perform kMC simulations with both the NI and FI model and evaluate short range order (SRO) parameters and cluster distributions. Using a spin-like variable $S_{i}$ (with $S_{i}$ = +1 and -1 for Ti and Ta, respectively) the Warren-Cowley SRO parameter can be expressed as a normalized pair-correlation function. For a binary alloy, the SRO parameter of site $i$ for the $n$th neighbor shell is defined as~\cite{sro-cite}
\begin{equation}\label{eq:sro}
    \text{SRO}_{n} = \frac{\langle S_{i}S_{i+n} \rangle - \langle S_{i} \rangle^2}{1 - \langle S_{i} \rangle^2}  \quad .
\end{equation}
Here, $\langle S_{i} \rangle$ = 2$x$ - 1 and $x$ represents the overall fraction of the chemical species occupying site $i$. $\langle$\ldots$\rangle$ refer to an ensemble average taken over all atoms of a given configuration. 
Taking as an example the chemically ordered B2 structures at 50\% Ta, SRO$_{n}$ takes a value of -1 for $n$ = 1, and +1 for $n$ = 2 and 3.  For a disordered random alloy, the value of SRO$_{n}$ would be 0 for all $n$. According to the definition in Eq.~\eqref{eq:sro}, a positive value of SRO$_n$ for $n$ = 1 and $\langle S_{i} \rangle$ = 0 (for 50\% Ta composition) implies that atoms of the same type (either Ti or Ta) prefer to form clusters. We calculate SRO$_{n}$ ($n$ = 1, 2 and 3) for Ti-25\% Ta, Ti-50\%Ta, and Ti-75\% Ta alloys within the NI and FI models for $T$ = 900~K. We choose $T$ = 900~K as the temperature to investigate  the transport properties where Ti-Ta alloys behave as a solid-solution~\cite{Murray-expt, Murray-theo, BARZILAI2016255} over almost the entire composition range.
The NI model serves here as a reference for a completely random alloy.
For Ti-25\% Ta and Ti-75\% Ta, both the NI and FI model result in random solid solutions, no chemical ordering or clustering is observed.
For Ti-50\% Ta, small differences between the results obtained within the NI and FI models are found.  Since any difference are largest at this equiatomic composition, we discuss this case in detail.
In Fig.~\ref{fig-5}(a) we present  SRO$_{1}$ values calculated for Ti-50\% Ta within the NI and FI model. 
As expected, SRO$_{1}$ fluctuates around zero for the NI model, signifying a random solid solution.  Within the FI model, some clustering is indicated by SRO$_{1} > 0$.  The value is, however, small and not sufficient to characterize any chemical order or phase separation.
The small amount of clustering is caused by the attractive first neighbor Ti-Ti interactions (cf. Fig.~\ref{fig-3}) incorporated in the FI model.
For SRO$_{2}$ and SRO$_{3}$, no significant differences between the results obtained within the NI and FI model are observed.

To further shed light on the formation of clusters predicted by the FI model, we carry out a cluster distribution analysis using a cluster algorithm in our kMC simulations. In the cluster algorithm, we define the minimum size of a cluster to consist of  9 Ti atoms. This represents the first-nearest neighbour shell around an atom in a bcc lattice.
The size of all clusters is monitored over 400 statistically independent configurations taken 10$^{5}$ kMC steps apart  to remove any correlations. The corresponding cluster size distribution is shown in Fig.~\ref{fig-5}(b).
The comparison of the cluster size distribution within the NI and FI model suggest, that although there are slightly more and slightly larger Ti clusters formed within the FI model, owing to the attractive Ti-Ti interactions, the differences are not significant enough to identify this as phase separation behavior. 

To quantify the clustering of Ti atoms further, we evaluate the \% of Ti atoms that participate in the cluster formation, shown in Fig.~\ref{fig-5}(c). From Fig.~\ref{fig-5}(c) we see that on average $\sim$~6.7\% and $\sim$~15.6\% of the Ti atoms are part of  the clusters  within the NI and FI model, respectively. These results again suggest, that although the attractive Ti-Ti interactions favor the formation clusters, the difference  as compared to a non-interacting random alloy is only minor. A full phase separation does not take place as only about $\sim$~15.6\% Ti atoms participate in the cluster formation. Overall, we find that the local interactions are not large enough to induce any ordering or phase separation and the Ti-Ta alloy remains a random solid solution over the entire composition range, also within the FI model.

\subsection{Transport properties}
\begin{figure*}
\begin{center}
\includegraphics[scale=0.22]{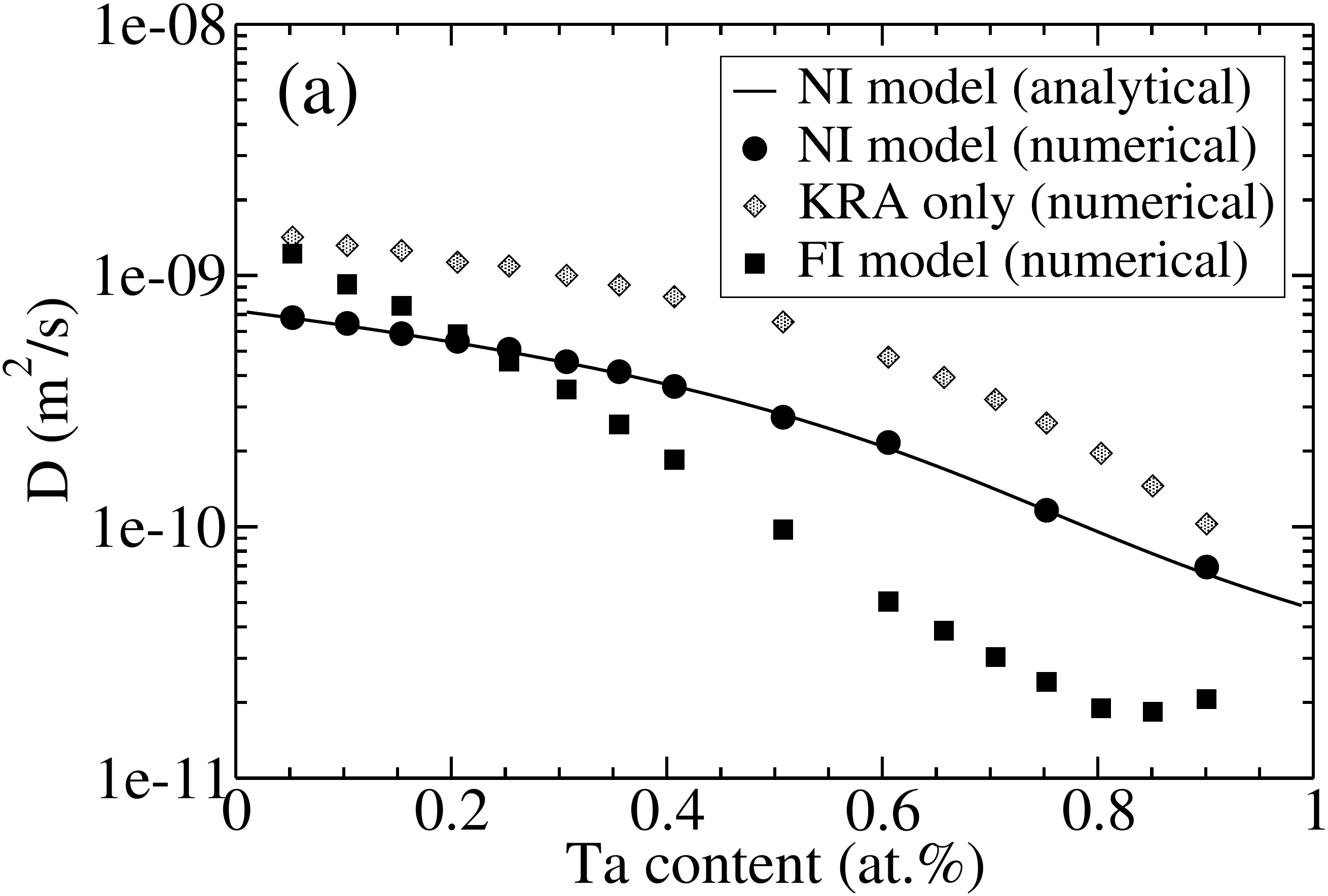}
\includegraphics[scale=0.22]{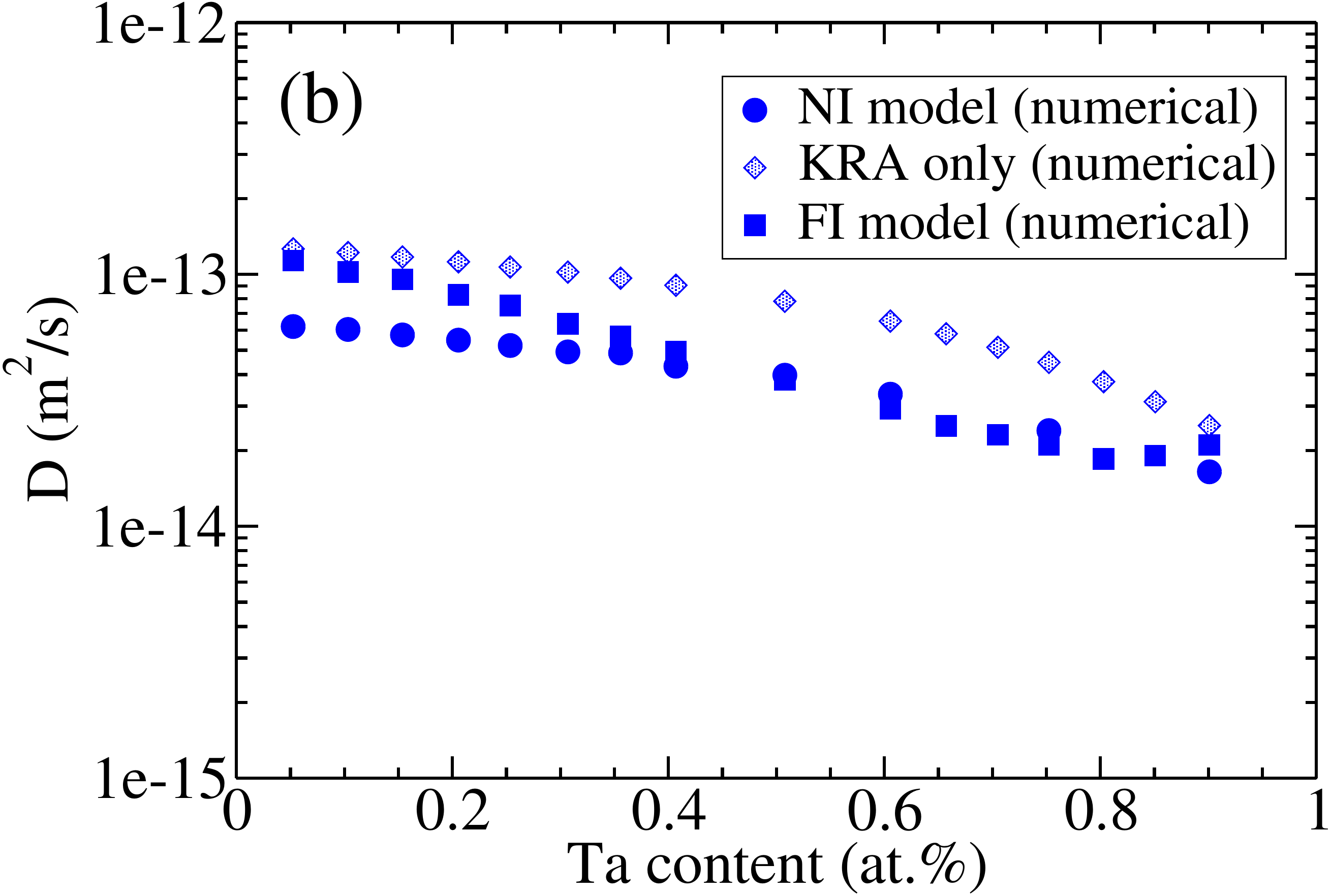}
\includegraphics[scale=0.22]{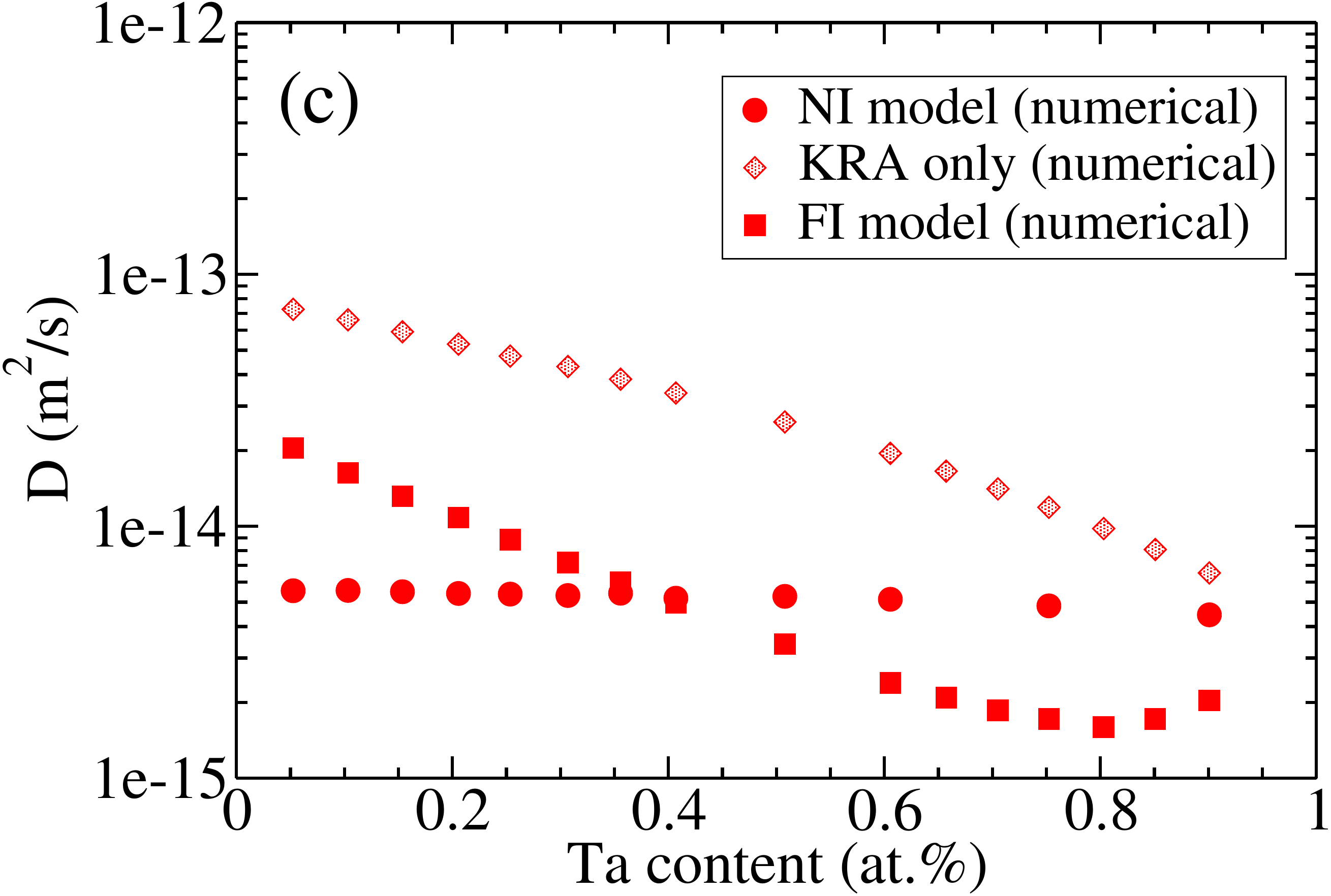}
\caption{\label{D}
(a) Vacancy diffusion coefficient computed with different diffusion models as a function of composition. The solid line and the circles represent the vacancy diffusivity within the NI model using analytical and numerical approaches, respectively. The diamonds are the data obtained from the KRA only model and the square symbols represent the values within the FI model. Tracer diffusion coefficient of (b) Ti and (c) Ta as a function of Ta content. The circle, diamond, and the square data points represent the values obtained within the NI, KRA only, and FI model, respectively. All values are calculated at $T = 900$~K.}
\end{center}
\end{figure*}

\subsubsection{Diffusivity}
Using the different diffusion models, the effect of local interactions on the transport properties is investigated as a function of composition. In Fig.~\ref{D}(a) we compile the vacancy diffusivity at $T$ = 900~K as a function of Ta content, extracted using both analytical and numerical approaches for the  different diffusion models. As expected, the analytical and numerical results within the NI model are in excellent agreement.  An overall  decrease in vacancy diffusivity with  increasing Ta content is observed. This is consistent with the general finding~\cite{VANDERVEN201061} that the vacancy diffusivity is reduced when the content of the slow diffusing species increases.

The FI model predicts a higher vacancy diffusivity  at low Ta content  compared to the NI model, which gradually slows down with increasing of Ta concentration and at $\sim$~80\% exhibits a {\it dip} followed by a slight increase in diffusivity. The results of the FI model is quite interesting and different from  the NI model, in particular the dip in mobility at around 80\% Ta. Overall, we find that including local interactions within the FI model for these alloys significantly impacts the vacancy diffusivity.

In order to further understand the origin of these differences in vacancy diffusivity, we use a rather hypothetical diffusion model which we call the {\it KRA only} model. In the KRA only model, we assume that the migration barrier height may change due to the local chemical environment, but the potential energy landscape remains flat. This implies that diffusion barriers are always symmetric, i.e., the local interactions do not influence the energy of the end-points of the diffusion process and, thus, $\Delta E = 0$. In other words, in the KRA only model the diffusion barriers resemble the average of the forward and backward diffusion barriers of the corresponding FI model. 
In our kMC simulations, we still utilize Eqs.~\eqref{Ti_fit} and~\eqref{Ta_fit} but set $\Delta E = 0$ in Eq.~\eqref{eq:DE}, correspondingly. From the results within the KRA only model shown in Fig.~\ref{D}(a) we see that a similar trend for the vacancy diffusivity is observed as within the NI model, where the mobility tends to decrease with increasing Ta content. However, the absolute values of the vacancy diffusivity are somewhat overestimated, especially at $\sim$~50\% Ta content compared to both the NI and FI model. Moreover, we do not observe a dip as found within the FI model. 
The results obtained within the KRA only model suggest that the dependence of only the barrier height on the local chemical environment is not sufficient to account even for the trends observed within the FI model.  It seems that the ruggedness of the energy landscape, i.e. the difference in energy between different configurations, is essential to capture the change in diffusion behavior induced by the local interactions.

Fig.~\ref{D}(b) and Fig.~\ref{D}(c) represent the tracer diffusion coefficients  of Ti and Ta as a function of Ta content, respectively, computed using the different diffusion models. The overall trends obtained within the different diffusion models is very similar to the ones for the vacancy diffusivity. For Ti tracer diffusion, we observe again that at low Ta content the FI and KRA only model predict a faster diffusion than the NI model which suggests that local interactions facilitate the diffusion of Ti atoms. For Ta tracer diffusion, the differences among the different diffusion models is very prominent. We find that the results obtained from the KRA only model is largely overestimating the diffusion coefficient over the entire composition range compared to both the Ni and FI models. The {\it dip} observed at $\sim$ 80\% Ta content within the FI model, which none of the other two models can reproduce, is more subtle for Ta than for Ti.

The vacancy diffusivity and tracer diffusion coefficient in Ti-Ta obtained within the different models suggest that even for a seemingly simple, nearly random alloys simplified diffusion models are not sufficient to accurately estimate the composition dependent diffusion coefficients of the involved species. A more involved diffusion model that properly includes local interactions is thus necessary.

\subsubsection{Onsager transport coefficients}
\begin{figure}
\begin{center}
\includegraphics[scale=0.32]{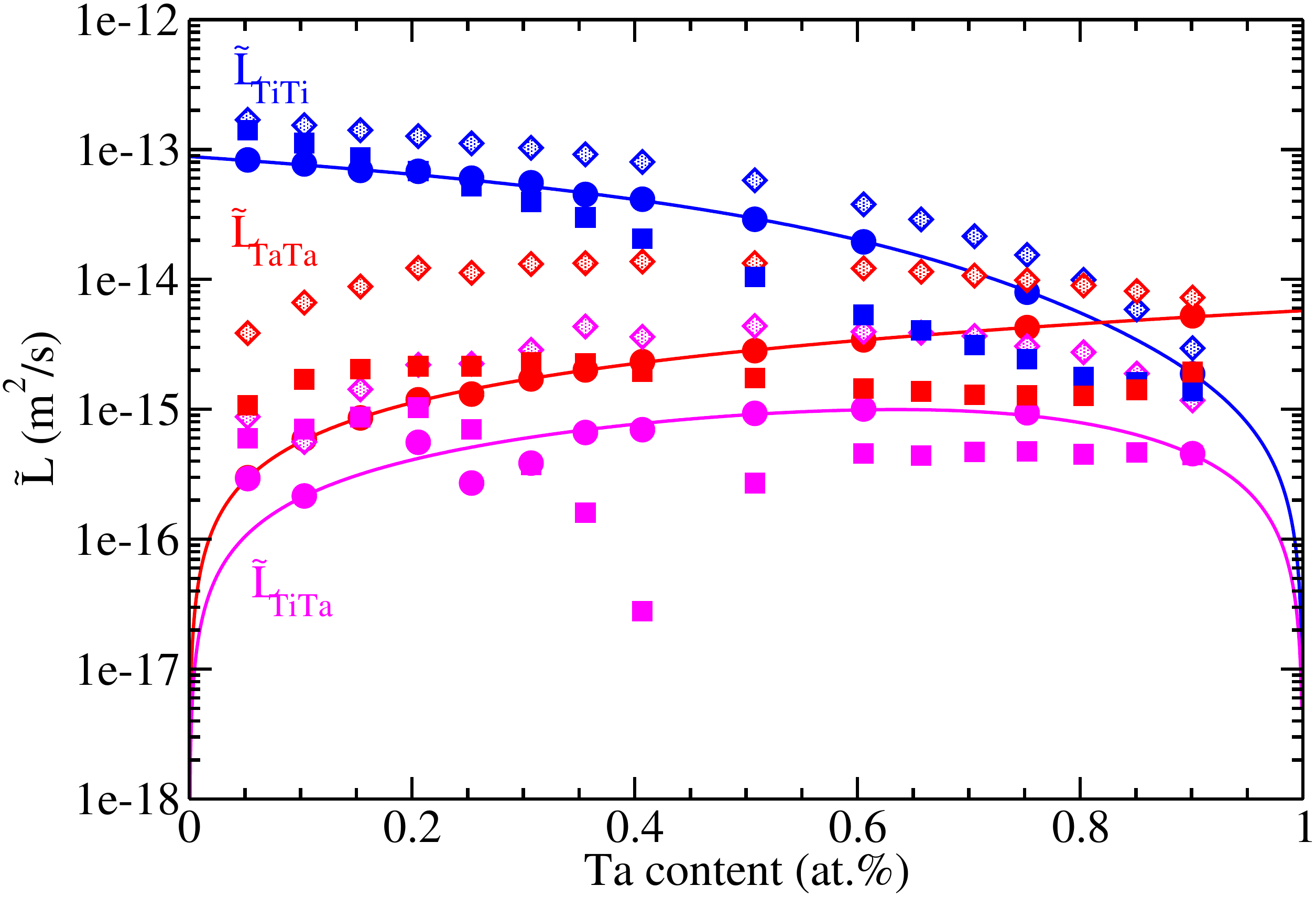}
\caption{\label{L}
The diagonal and off-diagonal Onsager transport coefficients as a function of Ta content computed within the NI, KRA only, and FI models at $T$ = 900 K. The blue, red, and magenta color represent  $\tilde{L}_{\text{TiTi}}$, $\tilde{L}_{\text{TaTa}}$, and $\tilde{L}_{\text{TiTa}}$, respectively. The solid lines represent the analytical solution within the NI model. The circle, diamonds, and the square symbols represent the values computed within the numerical NI, KRA only, and FI models, respectively. $\tilde{L}_{\text{TiTa}}$ obtained from the FI model, above 41\% Ta content, represent absolute values.}
\end{center}
\end{figure}
We also investigate the impact of local interactions on Onsager phenomenological transport coefficients as they provide important physical information about the nature of the involved chemical species and vacancy fluxes. Since these coefficients are related to the diffusion coefficients, they are often calculated along with diffusion coefficients in the literature~\cite{PhysRevB.96.094108, Goiri, VANDERVEN201061}. Fig.~\ref{L} shows Onsager transport coefficients calculated within the NI, KRA only, and FI models over the entire composition range at $T = 900$~K. The results of the analytical and numerical solution within the NI model are again in excellent agreement for all  three transport coefficients ($\tilde{L}_{\text{TiTi}}$, $\tilde{L}_{\text{TaTa}}$, and $\tilde{L}_{\text{TiTa}}$). The numerical results within the FI model are again quite different compared to the NI model, especially at very low and high Ta content, similar to what we have observed for the diffusivities. At low Ta content, the values of all  three Onsager transport coefficients are higher within the FI model than within the NI model, and  the diagonal transport coefficients ($\tilde{L}_{\text{TiTi}}$, $\tilde{L}_{\text{TaTa}}$) show a {\it dip} at $\sim$ 80\% Ta content. The cross term $\tilde{L}_{\text{TiTa}}$, computed within the FI model, becomes negative above 41\% Ta content. Thus, above 41\% Ta, the absolute values of $\tilde{L}_{\text{TiTa}}$ are  shown in Fig.~\ref{L}. This observation is similar to what Goiri $\emph{et al.}$ found for Ni-Al alloys~\cite{Goiri} where $\tilde{L}_{\text{NiAl}}$ becomes negative below 78\% Ni content. The transport coefficients within the KRA only model are overestimated compared to both the NI and FI models and also  the dip in $\tilde{L}_{\text{TiTi}}$ and $\tilde{L}_{\text{TaTa}}$ at $\sim$ 80\% Ta observed within the FI model is not captured. Overall, we see again the importance of local interactions that ultimately affect the macroscopic transport properties notably.
\subsection{Correlation factor}\label{sec:correlation}
\begin{figure*}
\begin{center}
\includegraphics[scale=0.23]{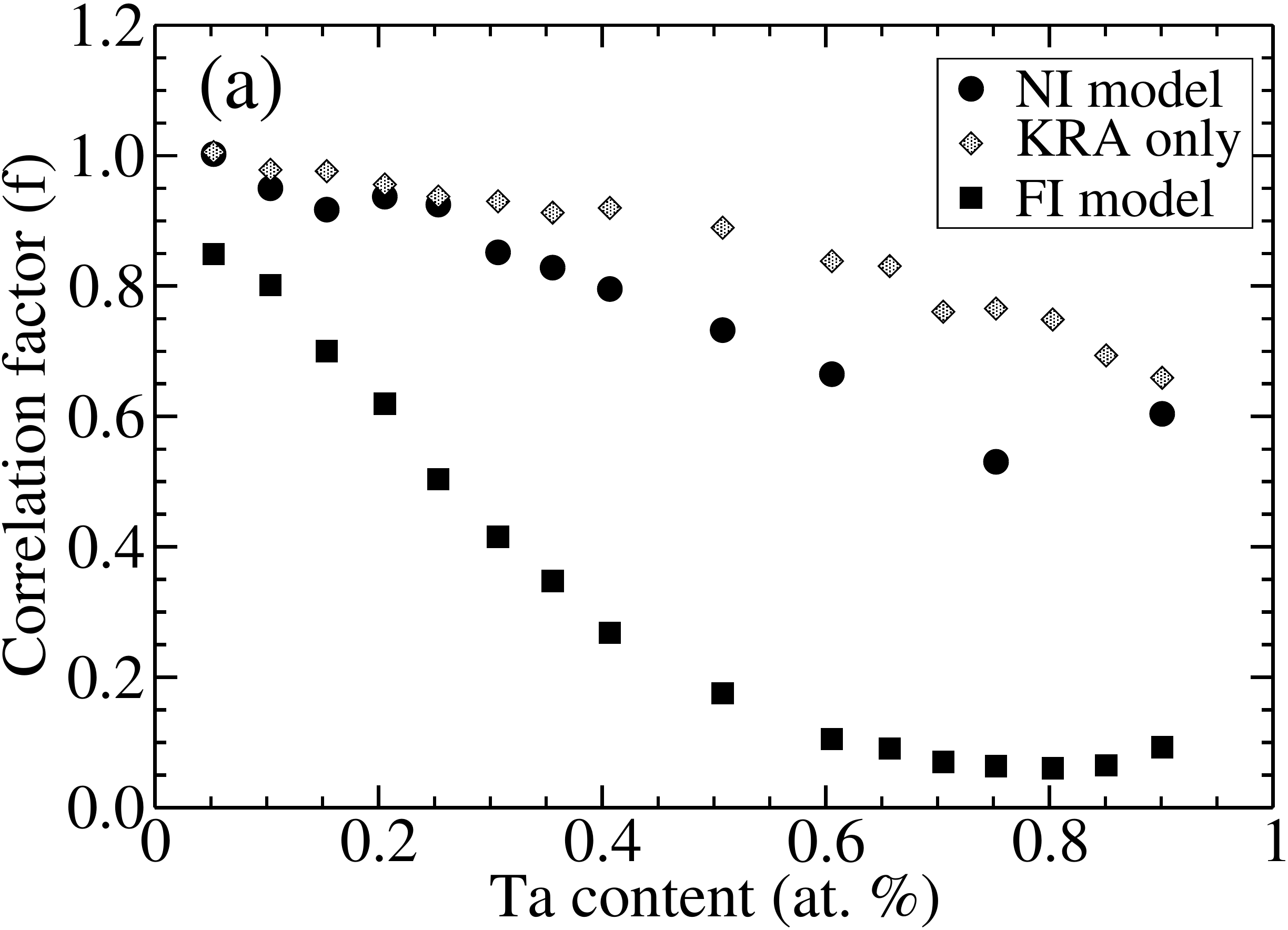}
\includegraphics[scale=0.23]{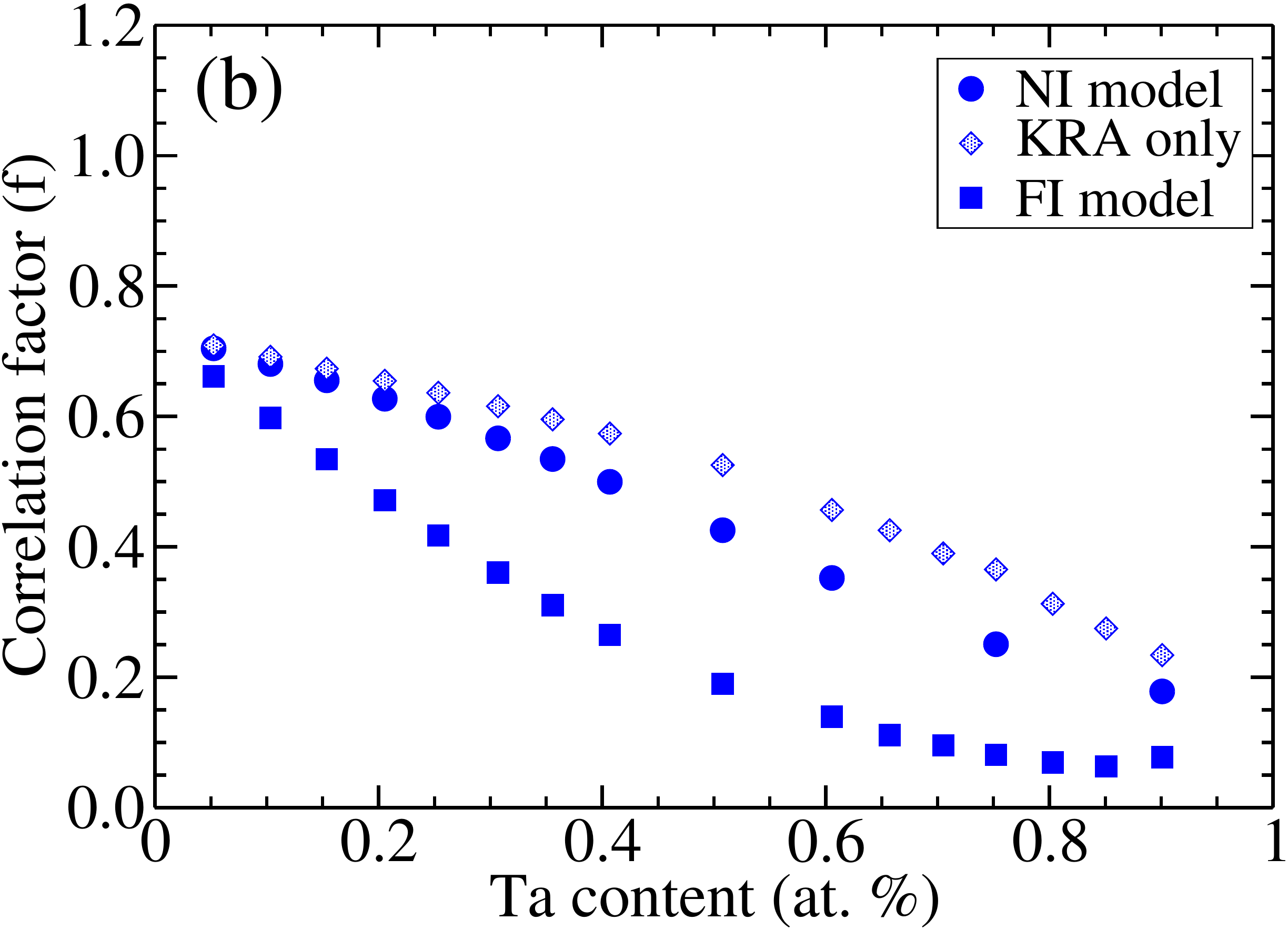}
\includegraphics[scale=0.23]{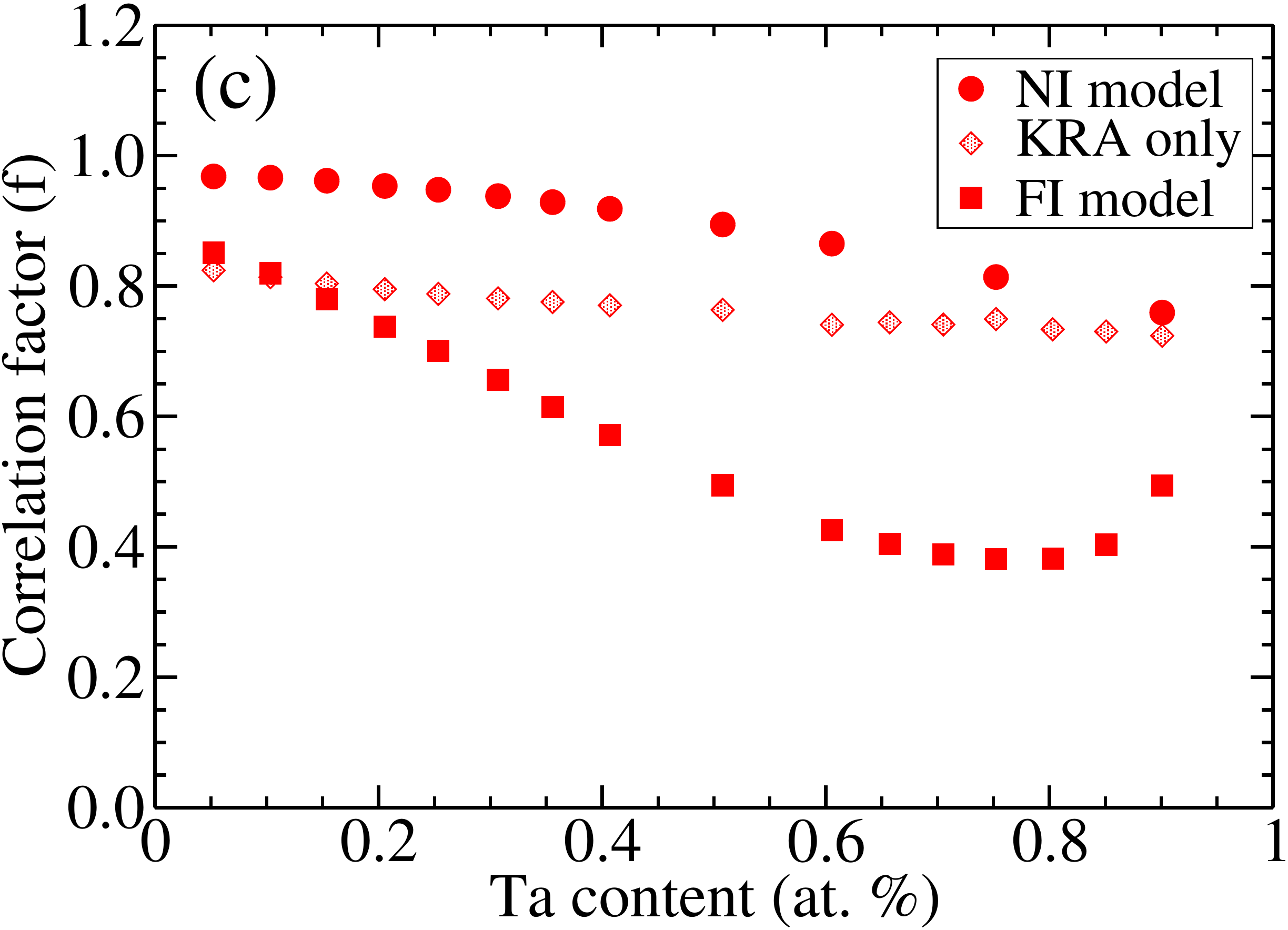}
\caption{\label{f}
Correlation factors as a function of Ta content for (a) vacancy, (b) Ti, and (c) Ta using the NI (circles), KRA only (diamonds), and FI (squares) models calculated at $T = 900$~K. 
}
\end{center}
\end{figure*}
The correlation factor, $f$, provides useful information about the role of short- and long-range order on diffusion mechanisms which can be defined for a species $i$ as 
\begin{equation}\label{eq:f}
    f_{i} = \frac{\langle \Delta \Vec{R}^{2}_{i} \rangle}{na^{2}} \quad ,
\end{equation}
where $\Delta \Vec{R}^{2}_{i}$ denotes the endpoints of the trajectory of a particle after $n$ hops and $a$ represents the elementary hop distance e.g. the nearest-neighbor distance. The brackets again denote an ensemble average in thermodynamic equilibrium. A value of $f=1$ represents a completely random walk. Any deviation from $f=1$ implies that the motion is correlated. The calculated correlation factors for the vacancy, Ti, and Ta are shown in Fig.~\ref{f} as a function of composition within the NI, KRA only, and FI models. Overall, the correlation factor decreases with  increasing  Ta content for all  diffusion models. However, the value of the correlation factor for Ti, Ta, and vacancy obtained within the FI model is significantly lower than within the other models for a given Ta content. At $\sim$~80\% Ta, the correlation factors of the vacancy, Ti, and Ta exhibit a {\it dip}, similar as the behavior observed for the diffusivities. This suggests that the change in mobility of the involved species in these alloys is strongly affected by a change in the correlation factor and not so strongly by the change in barriers. The differences in the correlation factors within the different models indicate that the local interaction energies can produce significant correlated motion which in turn affect the diffusion properties in these alloys.

\section{Summary and conclusion}\label{Summary and conclusion}
We have carried out a comprehensive and systematic first-principles multi-scale study of diffusion properties in Ti-Ta nearly random binary alloys over the entire composition range. Using a DFT parametrized CE Hamiltonian, we incorporate local interactions in these alloys which are predominantly attractive Ti-Va and Ti-Ti pair interactions. Subsequently, kMC simulations are performed using these CE Hamiltonians to compute macroscopic transport properties of the vacancy and Ti and Ta atoms. 

We observe significant deviations in transport properties  upon incorporating local interactions, as compared to a non-interacting solid solution model. The analysis of SRO parameters and cluster distributions, however, suggests that the local interactions are not strong enough to induce ordering or phase separation  in these alloys, but the system  remains a random solid solution over the entire composition range at the given temperature. Furthermore, the outcome of a rather hypothetical {\it KRA only} model, where the potential energy landscape is \emph{flat} and the local interactions only affect the KRA barriers turns out to be significantly over-estimating the diffusion properties compared to the NI and FI models. 
The KRA only model also fails to predict the correct trend of the composition-dependent macroscopic transport properties as compared to the FI model. 

Based on our calculations, we demonstrate that the changes in mobility  are predominantly connected to changes in the correlation factor caused by the local interactions in these alloys. Simpler diffusion models, such as the NI and KRA only models, are insufficient to accurately estimate the transport properties in these alloys, even though the thermodynamic properties
suggest that these systems still behave like random alloys. Our work thus highlights the importance of local interactions on transport properties in non-dilute,  thermodynamically nearly disordered binary alloys.

\begin{acknowledgments}
The work presented in this paper has been financially supported by the Deutsche Forschungsgemeinschaft (DFG) within the research unit FOR 1766 (High Temperature Shape Memory Alloys, http://www.for1766.de, Project No. 200999873, sub-project 3) and the DFG Heisenberg Programme  Project No. 428315600. The authors thank Dallas R. Trinkle for helpful and inspiring discussions.
\end{acknowledgments}

\appendix

\section{Expressions of $\Lambda$ and F}
\label{sec:appA}
The expressions of $\Lambda$ and $F$ are given by~\cite{Moleko, VANDERVEN201061}
\begin{multline}
    \Lambda = \frac{1}{2} (F+2)(x_{\text{A}}\Gamma_{\text{A}} + x_{\text{B}}\Gamma_{\text{B}}) - \Gamma_{\text{A}} - \Gamma_{\text{B}} + \\ 2(x_{\text{A}}\Gamma_{\text{B}} + x_{\text{B}}\Gamma_{\text{A}}) + \\
    \sqrt{\bigg(\frac{1}{2}(F+2)(x_{\text{A}}\Gamma_{\text{A}} + x_{\text{B}}\Gamma_{\text{B}}- \Gamma_{\text{A}} - \Gamma_{\text{B}})\bigg)^{2} + 2F\Gamma_{\text{A}}\Gamma_{\text{B}}}
\end{multline}
and 
\begin{equation}
    F = \frac{2f}{1-f} \quad ,
\end{equation}
where $f$ is the correlation factor of the given crystal structure.

\section{Cluster expansion formalism}\label{CE-formalism}
Within the CE formalism~\cite{CE1, CE2}, the total energy of a given configuration of an alloy can be expressed as
\begin{equation}\label{eq:CE}
    E(\vec{\sigma}) = V_{0} + \sum V_{\alpha} \phi_{\alpha} \quad , 
\end{equation}
where $\phi_{\alpha}$ is given by
\begin{equation}
    \phi_{\alpha} = \prod_{i \in \alpha} \sigma_{i} \quad .
\end{equation}
Here, $V_{\alpha}$ are the unknown parameters known as ``effective cluster interactions" (ECIs) and 
$V_{0}$ is a constant term representing the ECI for the \emph{empty} cluster. $\alpha$ denotes the size of the different cluster such as point, pair, triplet etc., and the sum runs over all the sites starting from 1.


\end{document}